\documentclass[journal]{IEEEtran}
\usepackage{graphicx}
\usepackage{amssymb}
\usepackage{amsmath}
\DeclareMathOperator{\Tr}{Tr}
\DeclareMathOperator{\vect2}{vec}
\DeclareMathOperator{\diag}{diag}
\DeclareMathOperator{\Diag}{Diag}
\usepackage{algorithmic}
\usepackage{array}
\usepackage[caption=false,font=footnotesize]{subfig}
\usepackage{stfloats}
\usepackage{url}
\usepackage[short]{optidef}
\usepackage{bm}
\usepackage{makecell}
\usepackage{algorithmic}
\usepackage{booktabs}
\usepackage{adjustbox}
\usepackage{multirow}
\usepackage{cite}
\usepackage[linesnumbered,ruled,vlined]{algorithm2e}
\usepackage{setspace}
\def\BibTeX{{\rm B\kern-.05em{\sc i\kern-.025em b}\kern-.08em
    T\kern-.1667em\lower.7ex\hbox{E}\kern-.125emX}}

\SetKwInput{KwInput}{Input}                
\SetKwInput{KwOutput}{Output}              
\ifCLASSINFOpdf
\else
\fi

\begin{document}
\title{Rate-Splitting Multiple Access for Multi-Antenna Joint Radar and Communications with Partial CSIT: Precoder Optimization and Link-Level Simulations}

\author{Rafael~Cerna~Loli,~\IEEEmembership{Student~Member,~IEEE},
        Onur~Dizdar,~\IEEEmembership{Member,~IEEE},
        and~Bruno~Clerckx,~\IEEEmembership{Fellow,~IEEE}
\thanks{This paper was presented in part at 22nd IEEE International Workshop on Signal Processing Advances in
Wireless Communications (SPAWC), Lucca, Italy, September 2021.

R. Cerna, O. Dizdar and B. Clerckx are with Imperial College London, London SW7
2AZ, UK (email: rafael.cerna-loli19@imperial.ac.uk;  o.dizdar@imperial.ac.uk; b.clerckx@imperial.ac.uk).}}


\maketitle

\begin{abstract}
Dual-Functional Radar-Communication (DFRC) systems have been investigated to manage the inter-system interference between radar and communication systems. However, the studies in literature often assume that the DFRC possesses perfect Channel State Information at the Transmitter (CSIT), which is an unrealistic assumption due to the inevitable CSIT errors in practical deployments. In this work, we aim to design a DFRC system under the practical assumption of partial CSIT. To achieve this, the proposed DFRC marries the capabilities of a Multiple-Input Multiple-Output (MIMO) radar with Rate-Splitting Multiple Access (RSMA). RSMA is a powerful downlink communication scheme based on linearly precoded Rate-Splitting (RS) that partially decodes multi-user interference (MUI) and partially treats it as noise and is inherently robust to partial CSIT. Using RSMA, the DFRC precoders are optimized in the presence of partial CSIT to simultaneously maximize the Average Weighted Sum-Rate (AWSR) under Quality-of-Service (QoS) constraints and minimize the DFRC beampattern Mean Squared Error (MSE) against an ideal MIMO radar beampattern. Simulation results demonstrate that the RSMA-based DFRC largely outperforms DFRCs based on other commonly used strategies such as Space Division Multiple Access (SDMA) and Non-Orthogonal Multiple Access (NOMA). Specifically, the common stream unique in the RSMA-based DFRC allows for flexible rate partitioning to guarantee user rate fairness with partial CSIT while also being the main contributor to generating a directional beampattern for effective radar sensing. The practical performance of the DFRC is then further assessed through Link-Level simulations (LLS) to take into account the effects of coding and modulation in the finite length regime as well as the channel aging due to mobility and latency, where the superiority of RSMA is again corroborated.
\end{abstract}


\IEEEpeerreviewmaketitle

\section{Introduction}
 With the introduction of next generation wireless systems, a need arises to efficiently manage and allocate the limited resources of the electromagnetic (EM) spectrum without introducing high interference levels to currently deployed wireless systems. Such is the case of the increasing EM spectrum congestion in the sub-10 GHz bands \cite{ucl_crss, crss_data_2}, where radar systems, crucial for public safety and military applications, and commercial wireless communication systems, such as the new 5G-NR mobile communication networks, Internet of Things (IoT) networks and the already existing LTE mobile communication networks, are competing for the already scarce EM resources. Therefore, two questions are raised:
\begin{itemize}
    \item How should the EM spectrum be allocated so that radar and communication systems can operate properly and simultaneously on the same frequency band?
    \item How can the interference between radar and communication systems be efficiently managed?
\end{itemize}

These challenges are the focus of the research area named Communication and Radar Spectrum Sharing (CRSS). CRSS research is aimed to be applied not only in military and commercial mobile communications applications, but also, for instance, in Vehicle-to-Everything (V2X) networks, WiFi localization, Unmanned Aerial Vehicle (UAV) networks, medical sensors and radar relay applications \cite{ucl_crss}. Several efforts have been made to optimize different performance metrics of spectrum sharing radar and communication systems by employing techniques such as interference mitigation, beamforming, and optimum waveform design. Nonetheless, research can generally be classified into three categories: coexistent, cooperative and dual functional radar-communication systems \cite{radcom_survey,radcom_overview,crss_app}. 

A review of the first two system types, in which radar and communication systems with separate hardware are considered, has already been thoroughly given in the cited references. Thus, for our purposes it suffices to state that the coexistent and cooperative categories represent a short-term solution to the EM congestion problem and, thus, only seek to avoid or control the interference between radar and communications. On the other hand, dual-functional radar-communication (DFRC) systems represent a true unification of the hardware and signal processing units of radar and communication systems. Thus, the challenges of synchronization, hardware costs and privacy can be overcome compared to the coexistent and cooperative radar-communication systems. This is also the most suitable approach in the long-term development of wireless systems and EM spectrum allocation as it provides significant advantages such as  highly-directional beamforming, minimum delay, enhanced security and privacy, and dynamic computational resource allocation \cite{radcom_overview}.

Research in this category has attracted a large attention in recent years. For instance, a DFRC waveform is designed in \cite{fliu_mimo_radar}, where the communications precoder matrix is optimized so that the obtained beampattern matches a desired radar beampattern under Signal-to-Interference-and-Noise Ratio (SINR) constraints for each of the communication users using manifold algorithms. The DFRC waveform is also optimized under a constant-modulus constraint in order to approximate a desired radar beampattern and minimize the Multi-User Interference (MUI) among communication users in \cite{cm_radcom} and maximize the SINR of the communication users in \cite{cm_radcom_2}. However, it is necessary to highlight three limitations in the literature that hinder the analysis and practical performance evaluation of a DFRC. The first is the assumption of perfect Channel State Information at Transmitter (CSIT). Although unrealistic, this assumption is usually made to simplify the system design by being able to separate the user streams perfectly in the spatial domain in order to minimize the MUI between them. The second is the use of the radar beampattern approximation error as the main radar performance metric. While it is obvious that generating a highly directional beampattern in the direction of the radar target improves sensing performance, its relationship to other conventional and more concrete radar metrics, such as Radar Mutual Information (RMI) and Cramér-Rao Bound (CRB), is often not described. Finally, DFRC design has only been studied under idealistic conditions of Gaussian signalling, infinite constellation modulation size and infinite channel code length.

\begin{figure*}[t!]
		\centering
        \includegraphics[width=0.9\textwidth]{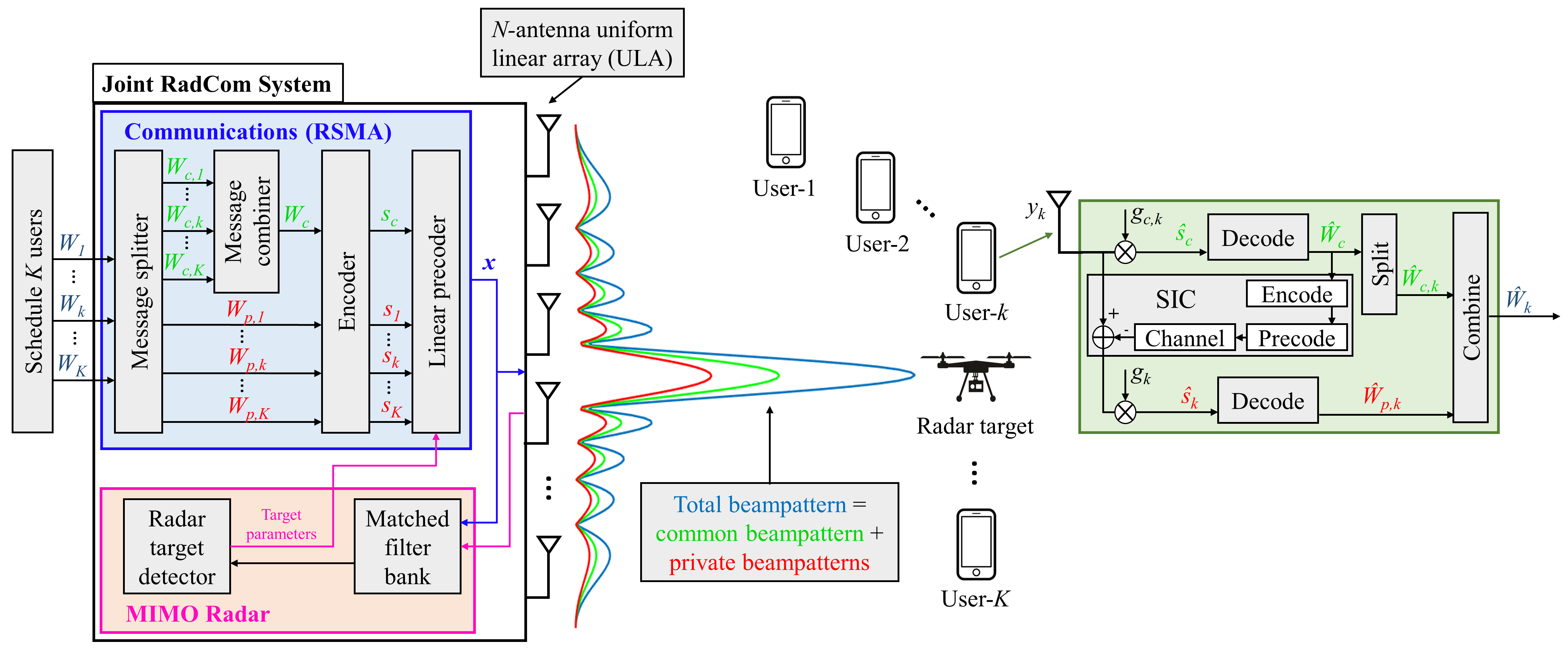}
		\caption{Proposed joint RadCom system model.}
		\label{fig:radcom_system_model}
\end{figure*}

 In this work, the benefits of employing Rate-Splitting Multiple Access (RSMA) to improve the performance of a DFRC in the presence of partial CSIT are explored. In order to achieve this, an RSMA communications module is considered to operate jointly with a Multiple-Input Multiple-Output (MIMO) radar module, as depicted in Fig. \ref{fig:radcom_system_model}. As it will be demonstrated in the following sections, RSMA constitutes a robust interference management framework for multi-antenna communication systems due to the splitting of user data streams into common and private streams \cite{rsma_lina}. In this way, all users decode the common streams (partially decoding MUI) and, after using Successive Interference Cancellation (SIC), each of them decodes its intended private stream while discarding the others (partially treating the remaining MUI as noise) \cite{rsma_clerckx}. Thus, RSMA can be seen as a multiple access strategy that bridges two conventional, yet extreme, multiple access schemes: Space Division Multiple Access (SDMA), which fully treats MUI as noise, and Non-Orthogonal Multiple Access (NOMA), which fully decodes MUI \cite{rsma_lina,perfect_1}. From an information theoretical perspective, this strategy translates into an increase in the number of Degrees of Freedom (DoFs), or interference-free streams, for each user \cite{joudeh}, \cite{joudeh_2} and, hence, an increase in the total Sum-Rate (SR) of the communication users. 
 
 Due to its great potential, RSMA has been widely analyzed in several applications with perfect CSIT \cite{rsma_lina,perfect_1,perfect_2,perfect_3,perfect_4,perfect_5} and partial CSIT \cite{rsma_clerckx,joudeh,joudeh_2}, \cite{imperfect_1,imperfect_2,imperfect_3,imperfect_4,imperfect_5,imperfect_6,imperfect_7,imperfect_8}. RSMA has also been studied in the context of DFRC design in \cite{chengcheng}, which considered an RSMA-based DFRC analysis with perfect CSIT for a specific channel realization, and in \cite{onur_radcom}, which proposed an RSMA-based DFRC with perfect CSIT and low resolution Digital-to-Analog Converter (DAC) units. The former demonstrated that the use of a separate radar sequence that is removed by a dedicated SIC layer is not necessary when using RSMA, while the latter showed that the RSMA-based DFRC outperforms the SDMA-based DFRC when using low resolution DACs regardless of the number of quantization bits. However, there are no previous works for the design of an RSMA-based DFRC with partial CSIT, in which the use of RSMA would offer a special advantage over other design procedures as its unique common stream can be employed to manage the unknown interference caused by CSIT inaccuracies while also guaranteeing the generation of a directional transmit beampattern that the radar module requires.

\subsection{Contributions}

The main contributions are then summarized as follows:

\textit{First}, this paper follows and extends some earlier work in \cite{chengcheng}, in which perfect CSIT allowed the RSMA-based DFRC to exclusively use the common stream to manage interference in azimuth directions in which the communication users experienced good channel conditions simultaneously. However, this is not possible with the uncertainty introduced by partial CSIT. Therefore, this work focuses on the more general design of an RSMA-based DFRC with partial CSIT and is the \textit{first paper} to study its ergodic sum-rate performance. Also, a model to consider the mobility of the communication users is included as a practical application.

\textit{Second}, the DFRC precoders are designed to simultaneously maximize the communication users Average Weighted Sum Rate (AWSR) under QoS constraints for different degrees of CSIT inaccuracies, and minimize the transmit beampattern Mean Squared Error (MSE), against a desired highly-directional radar beampattern. These two types of measurements are then jointly used as the main system performance metrics. In order to achieve this, a flexible precoder optimization algorithm based on the Alternating Direction Method of Multipliers (ADMM) is introduced. The main non-convex precoder optimization problem is divided into an AWSR maximization sub-problem and a MSE minimization sub-problem. The first sub-problem is then reformulated and solved via the Sample Average Approximation (SAA) and Weighted Minimum-Mean Square Error (WMMSE) method \cite{joudeh}. The second sub-problem is solved via Semi-definite Relaxation (SDR).

\textit{Third}, the performance of RSMA is compared with SDMA and NOMA when applied in the DFRC. Results demonstrate that employing RSMA provides the best results, for both radar and communications, as it proves to be more resilient against CSIT errors while also providing flexibility to comply with the QoS rate constraints and to approximate the desired directional radar beampattern.

\textit{Fourth}, the relationship between the beampattern MSE and the RMI and CRB is established in order to demonstrate that minimizing the radar beampattern MSE also maximizes the RMI and minimizes the CRB of the DFRC. Results confirm that employing the RSMA-based DFRC also improves the radar performance according to these other two more conventional radar metrics.

\textit{Fifth}, this paper presents the first Link-Level Simulations (LLS) for a DFRC, in order to assess its practical performance using finite-length polar codes and finite size QAM constellations. LLS results corroborate the superiority of RSMA for joint radar sensing and communications while also highlighting the challenges and losses incurred due to finite length modulation and coding schemes.  

\par \textit{Organization:}
The rest of the paper is organized as follows. The DFRC model is described in Section \ref{system_model_section} and the problem formulation is detailed in Section \ref{problem_section}. Then, the precoder optimization algorithm is explained in Section \ref{optimization_section}. Numerical and LLS results are presented in Sections \ref{numerical_section}. Finally, the conclusion is given in Section \ref{conclusion_section}.

\par \textit{Notation:} Scalars, vectors and matrices are denoted by standard, bold lower and upper case letters, respectively. $\mathbb{R}$ and $\mathbb{C}$ denote the real and complex domains. The transpose and Hermitian transpose operators are represented by $(.)^T$ and $(.)^H$, respectively. $\odot$ is the Hadamard product. The expectation of a random variable is given by $\mathbb{E}\{.\}$, and $\Re\{.\}$ and $\Im\{.\}$ indicate the real and imaginary parts of a complex number. $||.||_2$ is the $l_2$-norm operator, $||.||_\textit{F}$ is the Frobenius norm operator, $\Tr$(.) is the matrix trace operator and $\vect2$(.) is the operator that vectorizes a matrix. Finally, the operator $\diag$(.) extracts the diagonal vector from a matrix, the operator $\Diag$(.) constructs a diagonal matrix from a vector, $\mathbf{I}$ denotes the identity matrix, $\mathtt{\sim}$ denotes "distributed as" and $\mathcal{CN}(0,\sigma^2)$ denotes a Circularly Symmetric Complex Gaussian (CSCG) distribution with zero mean and variance $\sigma^2$.

\section{DFRC Model}
\label{system_model_section}
We consider a multi-antenna DFRC that tracks multiple radar targets while serving $K$ downlink communication users, indexed by the set $\mathcal{K} = \{1,\dots,K\}$. The DFRC is equipped with $N$ antennas, arranged in an Uniform Linear Array (ULA) structure, which are used to both transmit the DFRC signal and receive the reflected echoes from the radar target. It is proposed that the DFRC utilizes an RSMA communications module and a mono-static MIMO radar, which are constantly exchanging key information about the communication users (i.e. transmit communication signals) and the radar targets (e.g. velocity, azimuth direction) to optimize the precoders jointly for both communication and radar functions. In the following subsections, the RSMA-based DFRC signal model, baselines SDMA-based DFRC and NOMA-based DFRC signal models, and channel model are described.

\subsection{RSMA-based DFRC Signal Model}
In the RSMA framework, the message of user-$k$ $W_k$ is split into a common part $W_{c,k}$ and a private part $W_{p,k}$, $\forall k\in\mathcal{K}$. Then, the common parts of all $K$ users $\{W_{c,1},\dots,W_{c,K}\}$ are jointly encoded into a single common stream $s_c$, while the private parts $\{W_{p,1},\dots,W_{p,K}\}$ are encoded separately into $K$ private streams $\{s_1,\dots,s_K\}$. The data streams are then linearly precoded using the precoder $\mathbf{P} = [\mathbf{p}_c,\mathbf{p}_1,\dots,\mathbf{p}_K] \in \mathbb{C}^{N \times (K+1)}$, where $\mathbf{p}_c$ is the common stream precoder and $\mathbf{p}_k$ is the private stream precoder for user-$k$. The transmitted signal $\mathbf{x} \in \mathbb{C}^{N \times 1}$, subject to the transmit power constraint $\mathbb{E}\{||\mathbf{x}||^2\}\leq P_t$, is then defined as follows:
\begin{equation}
    \mathbf{x} = \mathbf{P}\mathbf{s} = \mathbf{p}_cs_c + \sum_{k=1}^{K}\mathbf{p}_ks_k,
    \label{rsma_transmit_signal}
\end{equation}
where $\mathbf{s} = [s_c,s_1,\dots,s_K]^T \in \mathbb{C}^{(K+1)\times 1}$. We assume that $\mathbb{E}\{\mathbf{s}\mathbf{s}^H\}=\mathbf{I}_{(K+1)}$ and $\Tr(\mathbf{P}\mathbf{P}^H)\leq P_t$.

We propose that the communication signal in (\ref{rsma_transmit_signal}) is also used for MIMO radar sensing. One of the key characteristics of a MIMO radar is its capacity to employ multiple probing signals with varying degrees of correlation between them \cite{convex_radar}. Based on this property, it has been shown in \cite{convex_radar}, \cite{stoica_radar} that the detection capabilities of a MIMO radar can be improved by appropriately designing the covariance matrix $\mathbf{R}_{\mathbf{x}} \in \mathbb{C}^{N \times N}$ of the transmitted signal $\mathbf{x}$. It has been further demonstrated in \cite{friedlander_radar} that optimum design of $\mathbf{R}_{\mathbf{x}}$ can be achieved in a simplified manner by generating $\mathbf{x}$ as a linear combination of independent signals, which agrees with the signal model in (\ref{rsma_transmit_signal}). In this way, optimization of $\mathbf{R}_{\mathbf{x}}$ can be effectively reduced to optimization of the precoder matrix $\mathbf{P}$. 

At user-$k$, the received signal is given by
\begin{equation}
    \begin{split}
        y_k &= \mathbf{h}_k^H\mathbf{P}\mathbf{s} + n_k \\
        &= \mathbf{h}_k^H\mathbf{p}_c s_c+ \mathbf{h}_k^H\mathbf{p}_k s_k+\overbrace{\sum_{j\neq k,j\in\mathcal{K}}\mathbf{h}_k^H\mathbf{p}_j s_j}^{\text{MUI}} + n_k,
    \end{split}
    \label{rsma_eq}
\end{equation}
where $\mathbf{h}_k \in \mathbb{C}^{N \times 1}$ is the channel between the DFRC and user-$k$, and $n_k\;\mathtt{\sim}\; \mathcal{CN}(0,\sigma_{n,k}^2)$ is the Additive White Gaussian Noise (AWGN) at user-$k$. Without loss of generality, we assume that $\sigma_{n,k}^2 = \sigma_n^2 = 1,\; \forall k \in \mathcal{K}$. First, the common stream $s_c$ is decoded into $\hat{W}_c$ by treating the interference from the $K$ private streams as noise. Thus, SINR of decoding the $s_c$ is given by
\begin{equation}
    \gamma_{c,k} = \frac{|\mathbf{h}_k^H\mathbf{p}_c|^2}{\sum_{k\in\mathcal{K}}|\mathbf{h}_k^H\mathbf{p}_k|^2+\sigma_{n,k}^2}.
    \label{commonsinr}
\end{equation}
Then, the common stream is reconstructed using $\hat{W}_{c}$ and subtracted from $y_k$ using SIC so that the private stream $s_k$ can be decoded into $\hat{W}_{p,k}$ by treating the remaining $K-1$ private streams as noise. The SINR of decoding $s_k$ is then given by
\begin{equation}
    \gamma_k = \frac{|\mathbf{h}_k^H\mathbf{p}_k|^2}{\sum_{j\neq k,j\in\mathcal{K}}|\mathbf{h}_k^H\mathbf{p}_j|^2+\sigma_{n,k}^2}.
    \label{privatesinr}
\end{equation} 
Finally, user-$k$ extracts $\hat{W}_{c,k}$ from $\hat{W}_c$ and combines it with $\hat{W}_{p,k}$ to reconstruct the message $\hat{W}_k$. Thus, the achievable rate of the common stream for user-$k$, assuming Gaussian signalling, is  $R_{c,k}(\mathbf{P}) = \log_2(1+\gamma_{c,k})$ and the achievable rate of its corresponding private stream is $R_k(\mathbf{P}) = \log_2(1+\gamma_k)$. To guarantee that all $K$ users decode the common stream successfully, it must be transmitted at a rate that does not exceed $R_c(\mathbf{P}) = \min\{R_{c,1}(\mathbf{P}),\dots,R_{c,K}(\mathbf{P})\}$. We then have $\sum_{k\in\mathcal{K}}C_k=R_c(\mathbf{P})$, where $C_k$ denotes the portion of the common stream rate carrying $W_{c,k}$.  
    
\subsection{Baseline SDMA-based DFRC and NOMA-based DFRC Signal Models}
\label{baseline_subsection}
The baseline SDMA-based DFRC is enabled by not allocating any power to the common stream precoder in (\ref{rsma_transmit_signal}). In turn, the NOMA-based DFRC functions by encoding the $K$ user data streams in a superposed manner ordered according to their effective scalar channel strengths after precoding. Then, each user decodes and employs SIC to remove MUI from users with weaker channel strengths. Thus, for a given decoding order $\pi$, the data streams $s_{\pi(i)}$, $\forall i\leq k$ are decoded and removed using SIC at user-$\pi(k)$. The SINR of decoding $s_{\pi(i)}$, $i \leq k$ at user-$\pi(k)$ is then given by
\begin{equation}
    \gamma_{\pi(k) \rightarrow \pi(i)} = \frac{|\mathbf{h}_{\pi(k)}^H\mathbf{p}_{\pi(i)}|^2}{\sum_{j>i,j\in\mathcal{K}}|\mathbf{h}_{\pi(k)}^H\mathbf{p}_{\pi(j)}|^2+\sigma_{n,k}^2},
    \label{nomasinr}
\end{equation} 
where $\pi(j)$, $j>i, j\in\mathcal{K}$ is the decoding order of the stream $s_{\pi(j)}$, which has not been decoded at user-$\pi(k)$ at the moment of decoding $s_{\pi(i)}$.

\subsection{Channel Model}
In this work, we consider the following channel model
\begin{equation}
    \mathbf{H}=\hat{\mathbf{H}}+\Tilde{\mathbf{H}},
    \label{csi_eq}
\end{equation}
where $\mathbf{H}=[\mathbf{h}_1,\dots,\mathbf{h}_K]$ is the actual channel realization with the entries of $\mathbf{h}_k$ being i.i.d random variables with distribution $\mathcal{CN}(0,\sigma_k^2), \forall k\in \mathcal{K}$. $\hat{\mathbf{H}}=[\hat{\mathbf{h}}_1,\dots,\hat{\mathbf{h}}_K]$ is the CSIT with $\hat{\mathbf{h}}_k$ following a Gaussian distribution $\mathcal{CN}(0,\sigma_k^2-\sigma_{e,k}^2), \forall k\in \mathcal{K}$. Finally, $\Tilde{\mathbf{H}}=[\Tilde{\mathbf{h}}_1,\dots,\Tilde{\mathbf{h}}_K]$ represents the CSIT estimation error with $\Tilde{\mathbf{h}}_k$ following a Gaussian distribution $\mathcal{CN}(0,\sigma_{e,k}^2), \forall k\in \mathcal{K}$. We assume perfect Channel State Information at the Receiver (CSIR) and partial CSIT, which indicates that the DFRC knows only $\hat{\mathbf{H}}$ and its corresponding conditional CSIT error distribution $f_{\text{H}|\hat{\text{H}}}(\mathbf{H}|\hat{\mathbf{H}})$. Furthermore, we assume $\sigma_k^2 = 1, \forall k \in \mathcal{K}$, without any loss of generality.

\section{Performance Metrics and Problem Formulation}
\label{problem_section}
In this section, the DFRC performance metrics and optimization problem are described.

\subsection{Communications Metric: Average Weighted Sum-Rate}
Due to partial CSIT, calculation of the optimum precoders that maximize the sum of the common and private rates is not possible. A more robust approach with partial CSIT is then to optimize the precoders based on the Ergodic Rates (ERs) \cite{joudeh}. The common and private ERs of user-$k$ are then given by $\mathbb{E}_\text{H}\{R_{c,k}\}$ and  $\mathbb{E}_\text{H}\{R_{k}\}$, respectively. Moreover, $\min_k\{\mathbb{E}_\text{H}\{R_{c,k}\}\}_{k=1}^K$ denotes the common ER that guarantees successful decoding by all $K$ users.

To transform the stochastic problem of maximizing the ERs with only with partial CSIT into a deterministic one, we maximize the common and private Average Rates (ARs), short-term representations of the expected performance over the conditional error distribution $f_{\text{H}|\hat{\text{H}}}(\mathbf{H}|\hat{\mathbf{H}})$, over a sufficiently large number of random channel realizations $\hat{\mathbf{H}}$ \cite{joudeh}. The common and private ARs of user-$k$ are given by $\bar{R}_{c,k}\triangleq\mathbb{E}_{\text{H}|\hat{\text{H}}}\{R_{c,k}|\hat{\mathbf{H}}\}$ and  $\bar{R}_k\triangleq\mathbb{E}_{\text{H}|\hat{\text{H}}}\{R_{k}|\hat{\mathbf{H}}\}$, respectively. Additionally, the common AR for all $K$ users is denoted by $\bar{R}_c\triangleq\min_k\{\mathbb{E}_{\text{H}|\hat{\text{H}}}\{R_{c,k}|\hat{\mathbf{H}}\}\}_{k=1}^K$. The Average Weighted Sum-Rate (AWSR) metric is then defined as
\begin{equation}
    \text{AWSR}(\mathbf{P}) [\text{bps/Hz}] = \sum_{k\in\mathcal{K}}\mu_k(\bar{C}_k+\bar{R}_k(\mathbf{P})),
    \label{awsr_def}
\end{equation}
where $\mu_k$ is the weight assigned to user-$k$, and $\bar{C}_k\triangleq\mathbb{E}_{\text{H}|\hat{\text{H}}}\{C_{k}|\hat{\mathbf{H}}\}$.

\subsection{Radar Sensing Metric: Beampattern Mean Squared Error}
As discussed in \cite{fuhrmann_mimo_radar, fliu_mimo_radar,friedlander_radar}, the estimation capabilities of a MIMO radar are significantly improved by appropriately designing the covariance matrix $\mathbf{R}_{\mathbf{x}} \in \mathbb{C}^{N \times N}$ of the transmitted signal $\mathbf{x}$ to approximate a highly directional transmit beampattern $\bm{P}_d$ in the directions of the targets of interest. Thus, the radar sensing metric, the beampattern Mean Squared Error (MSE), can be defined as $ \sum_{m=1}^M|\alpha \bm{P}_d(\theta_m)-\mathbf{a}_t^H(\theta_m)\mathbf{R}_{\mathbf{x}}\mathbf{a}_t(\theta_m)|^2$, where $\alpha$ is the scaling factor of $\bm{P}_d$, $M$ is the total number of azimuth angle grids, $\theta_m$ is the $m$-th azimuth angle grid,  $\mathbf{a}_t(\theta_m) = [1,e^{j2\pi\delta\sin(\theta_m)},\dots,e^{j2\pi(N-1)\delta\sin(\theta_m)}]^T \in \mathbb{C}^{N \times 1}$ is the transmit antenna array steering vector at direction $\theta_m$ and $\delta$ is the normalized distance in units of wavelengths between antennas. For the proposed DFRC, the MSE is then given by
\begin{equation}
    \text{MSE}(\mathbf{P}) = \sum_{m=1}^M|\alpha \bm{P}_d(\theta_m)-\underbrace{\mathbf{a}_t^H(\theta_m)\mathbf{P}\mathbf{P}^H\mathbf{a}_t(\theta_m)}_{\bm{P}_t(\theta_m)}|^2,
    \label{mse_def}
\end{equation}
where $\bm{P}_t(\theta_m)$ is the DFRC transmit beampattern gain at direction $\theta_m$.

\subsection{Problem Formulation}
 For a given channel estimate $\hat{\mathbf{H}}$, the joint RadCom optimization problem with partial CSIT is formulated as
\begin{mini!}|s|[2]
  {\alpha,\bar{\mathbf{c}},{\mathbf{P}}}{\begin{aligned}[t]-\sum_{k\in\mathcal{K}}\mu_k(\bar{C}_k+\bar{R}_k({\mathbf{P}}))\qquad\qquad\qquad\qquad\;\;\;\\~\mathllap{+\lambda{\sum_{m=1}^M|\alpha \bm{P}_d(\theta_m)-\mathbf{a}_t^H(\theta_m)\big({\mathbf{P}}{\mathbf{P}}^H\big)\mathbf{a}_t(\theta_m)|^2}},\end{aligned}}
  {\label{main_design_awsr}}{}
  \addConstraint{\sum_{k'\in\mathcal{K}}\bar{C}_{k'}}{\leq \bar{R}_{c,k}({\mathbf{P}}),\quad\forall k \in \mathcal{K}\label{eq:C1Main_design_awsr}}{}
  \addConstraint{\bar{\mathbf{c}}}{\geq 0\label{eq:C2Main_design_awsr}}
  \addConstraint{\diag({\mathbf{P}}{\mathbf{P}}^H)}{=\frac{P_t\mathbf{1}}{N}\label{eq:C3Main_design_awsr}}
  \addConstraint{\alpha}{>0\label{eq:C4Main_design_awsr}}
  \addConstraint{(\bar{C}_k + \bar{R}_k({\mathbf{P}}))}{\geq \bar{R}_k^{th}\;,\;\forall k\in\mathcal{K},\label{eq:C5Main_design_awsr}}
\end{mini!}
\normalsize   
where $\bar{\mathbf{c}} = [\bar{C}_1,\dots,\bar{C}_K]^T \in \mathbb{R}_+^{K \times 1}$ is the variable vector that contains the portions of the common stream AR, $\bar{R}_c(\mathbf{P})$, allocated to the communication users, $\lambda$ is the regularization parameter to prioritize either communications (maximizing the AWSR) or radar sensing (minimizing the MSE), and $\bar{R}_k^{th}$ is the minimum total AR for user-$k$. Constraint (\ref{eq:C1Main_design_awsr}) ensures that $\bar{R}_c(\mathbf{P})$ is decodable by all $K$ users. Constraint (\ref{eq:C2Main_design_awsr}) forces the entries of $\bar{\mathbf{c}}$ to be positive for feasible partitioning of $\bar{R}_c(\mathbf{P})$ while constraint (\ref{eq:C3Main_design_awsr}) is introduced as a constant modulus constraint at each transmit antenna to avoid saturation of transmit power amplifiers in a practical scenario. Constraint (\ref{eq:C4Main_design_awsr}) forces $\alpha$ to be a positive scaling factor. Finally, constraint (\ref{eq:C5Main_design_awsr}) is an optional Quality of Service (QoS) rate constraint, which guarantees a minimum rate of $\bar{R}^{th}_{k}$ to be allocated to each user. 

\subsection{Other Conventional Radar Sensing Metrics}
\label{other_metric_disc}
The effectiveness of minimizing the MSE in (\ref{main_design_awsr}) can also be assessed by employing the optimized precoder matrix $\mathbf{P}$ to calculate other conventional radar sensing metrics, such as those described in the next subsections.
\subsubsection{Radar Mutual Information}
Assuming that the $(K+1)$ transmitted data streams are reflected on the single radar target in azimuth direction $\theta_0$, the effective frequency-domain radar sensing channel $\mathbf{H}_r \in \mathbb{C}^{N\times N}$ is given by
\begin{equation}
    \mathbf{H}_r= h_0e^{-j2\pi f_c(\tau_0-\frac{v_0}{c})}\mathbf{a}_t(\theta_0)\mathbf{a}^H_r(\theta_0),
    \label{path_loss_eq}
\end{equation}
where $h_0$ represents the path loss and target reflection amplitude, $\mathbf{a}_r(\theta_0) = [1,e^{j2\pi\delta\sin(\theta_0)},\dots,e^{j2\pi(N-1)\delta\sin(\theta_0)}]^T \in \mathbb{C}^{N \times 1}$ is the receive steering vector at direction $\theta_0$, $f_c$ is the carrier frequency, $\tau_0$ is the total time delay of the radar target, $v_0$ is the relative radar target velocity, and $c$ is the speed of light. The received signal at the DFRC is then given by 
\begin{equation}
    \mathbf{y}_r=\mathbf{H}_r^H\mathbf{P}\mathbf{s}+\mathbf{n}_r,
\end{equation}
where $\mathbf{n}_r\;\mathtt{\sim}\; \mathcal{CN}(0,\sigma_r^2\mathbf{I}_{N})$ is the AWGN.

\normalsize
The Radar Mutual Information (RMI) between $\mathbf{H}_r$, which contains all relevant parameters associated with the target of interest, and $\mathbf{y}_r$ can be calculated to measure radar performance as follows \cite{additional_metrics}

{\small
\begin{equation}
\begin{split}
    \text{RMI}(\mathbf{P}) &= I(\mathbf{y}_r;\mathbf{H}_r|\mathbf{s})=\log_2\Big(\Big|\mathbf{I}_{N}+\frac{\mathbf{H}_r^H\mathbf{P}\mathbf{P}^H\mathbf{H}_r}{\sigma_r^2}\Big|\Big)\\
   &=\log_2\Big(\Big|\mathbf{I}_{N}+\frac{|h_0|^2\mathbf{a}_r(\theta_0)\bm{P}_t(\theta_0)\mathbf{a}^H_r(\theta_0)}{\sigma_r^2}\Big|\Big).\\
\end{split}
\label{radar_mi_eq}
\end{equation}}

\normalsize
From (\ref{radar_mi_eq}), it is clear that maximizing the RMI is equivalent to maximizing the DFRC transmit beampattern gain $\bm{P}_t(\theta_0)$.

\normalsize
\subsubsection{Cramér-Rao Bound}
The Cramér-Rao Bound (CRB) can be used to give a theoretical lower bound on the estimation accuracy for the radar target parameters of interest. We then introduce the real-valued vector $\bm{\theta}=[\Re{\{h_0\}}, \Im{\{h_0\}}, \tau_0, v_0]$ to represent the parameters to be estimated (complex gain, velocity and range). The total CRB can then be calculated as $||\mathbf{B}||_{\textit{F}}=||\mathbf{F}^{-1}||_{\textit{F}}$, where $\mathbf{B}$ is the CRB matrix and $\mathbf{F}$ is the Fisher Information Matrix given by \cite{additional_metrics}

{\small
\begin{equation}
    \mathbf{F}=\frac{2}{\sigma_r}\begin{bmatrix}
    \Re(\mathbf{F}_1) & -\Im(\mathbf{F}_1) & \Re(\mathbf{F}_2) & \Re(\mathbf{F}3)\\ \Im(\mathbf{F}_1) & \Re(\mathbf{F}_1) & \Im(\mathbf{F}_2) & \Im(\mathbf{F}_3)\\
    \Re(\mathbf{F}2) & \Im(\mathbf{F}_2) & \Re(\mathbf{F}_4) & \Re(\mathbf{F}_5)\\
    \Re(\mathbf{F}_3) & \Im(\mathbf{F}3) & \Re(\mathbf{F}_5) & \Re(\mathbf{F}_6)
\end{bmatrix},
\label{crb_matrix}
\end{equation}}
\normalsize
and the elements $\mathbf{F}_1, \mathbf{F}_2, \mathbf{F}_3, \mathbf{F}_4, \mathbf{F}_5$ and $\mathbf{F}_6$ are given by

{\small
    \begin{align*}
        \mathbf{F}_1&=\mathbf{a}_r(\theta_0)\bm{P}_t(\theta_0)\mathbf{a}_r(\theta_0)^H, & \mathbf{F}_4&=-(2\pi f_ch_0)^2\mathbf{F}_1,\\
        \mathbf{F}_2&=-j2\pi f_ch_0 \mathbf{F}_1, & \mathbf{F}_5&=\frac{(2\pi f_ch_0)^2}{c} \mathbf{F}_1,\\
        \mathbf{F}_3&=j\frac{2\pi f_ch_0}{c} \mathbf{F}_1, & \mathbf{F}_6&=-\Big(\frac{2\pi f_ch_0}{c}\Big)^2 \mathbf{F}_1,\\
    \end{align*}
    \label{crb_defs}
}
\normalsize
where $c$ is the speed of the EM waves in air.

It can be observed that each of the elements in (\ref{crb_matrix}) are dependent of $\bm{P}_t(\theta_0)$. As also detailed in \cite{additional_metrics}, minimizing the total CRB is equivalent to minimizing the largest eigenvalue of $\mathbf{B}$. This, in turn, is equivalent to maximizing the smallest eigenvalue of $\mathbf{F}$, which is dependent of maximizing $\bm{P}(\theta_0)$ as it is a factor in all elements of $\mathbf{F}$.

\section{Precoder Optimization with Partial CSIT}
\label{optimization_section}
The optimization problem as defined in (\ref{main_design_awsr}) is difficult to directly solve due to its non-convex nature. Therefore, we propose an algorithm based on Alternating Direction Method of Multipliers (ADMM) \cite{admm} to obtain a solution for the considered problem. The proposed algorithm extends \cite{chengcheng}, which was proposed for DFRCs with perfect CSIT, for scenarios with partial CSIT. Before moving to describe it, we first introduce several parameters to help in our definitions.

The vector $\mathbf{v} = [\alpha, \bar{\mathbf{c}}^T, \vect2(\mathbf{P})^T]^T \in \mathbb{R}_{++} \times \mathbb{R}_+^{K \times 1} \times \mathbb{C}^{N(K+1) \times 1}$ is introduced to contain all optimization variables in (\ref{main_design_awsr}). We further define the selection matrices $\mathbf{D}_p = [\mathbf{0}^{(K+1)N\times (K+1)},\mathbf{I}_{(K+1)N}]$, $\mathbf{D}_c = [\mathbf{0}^{N \times (K+1)}, \mathbf{I}_{N}, \mathbf{0}^{N \times KN}]$ and $\mathbf{D}_k = [\mathbf{0}^{N \times (K+1+kN)}, \mathbf{I}_{N}, \mathbf{0}^{N \times (K-k)N}]$ $\forall k \in \mathcal{K}$, and selection vectors $\mathbf{f}_k = [\mathbf{0}^{1 \times k}, 1, \mathbf{0}^{1 \times [(K+1)N+K-k]}]^T$ $\forall k \in \mathcal{K}$.

The user ARs are expressed as $\bar{R}_{c,k}(\vect2(\mathbf{P}))=\bar{R}_{c,k}(\mathbf{D}_p\mathbf{v})$ and $\bar{R}_k(\vect2(\mathbf{P}))=\bar{R}_k(\mathbf{D}_p\mathbf{v})$. Then, (\ref{main_design_awsr}) is reformulated in an ADMM expression as follows
\begin{mini}
  {\mathbf{v},\mathbf{u}}{f_c(\mathbf{v})+g_c(\mathbf{v})+f_r(\mathbf{u})+g_r(\mathbf{u}),}{}{}
  \addConstraint{\mathbf{D}_p(\mathbf{v}-\mathbf{u})}{=0},
  \label{admm_design}
\end{mini}
where $\mathbf{u} \in \mathbb{R}_{++} \times \mathbb{R}_+^{K \times 1} \times \mathbb{C}^{N(K+1) \times 1}$ is a new optimization variable introduced in accordance with the ADMM framework. The functions $f_c(\mathbf{v})$ and $f_r(\mathbf{u})$ are defined as $f_c(\mathbf{v})=-\sum_{k\in\mathcal{K}}\mu_k(\mathbf{f}_k\mathbf{v}+\bar{R}_k\big(\mathbf{D}_p\mathbf{v}\big))$ and $f_r(\mathbf{u})=\lambda\sum_{m=1}^M|\alpha \bm{P}_d(\theta_m)-\mathbf{a}_t^H(\theta_m)\big(\mathbf{D}_c\mathbf{u}\mathbf{u}^H\mathbf{D}_c^H
+$\\ $\sum_{k\in\mathcal{K}}\mathbf{D}_k\mathbf{u}\mathbf{u}^H\mathbf{D}_k^H\big)\mathbf{a}_t(\theta_m)|^2$. $g_c(\mathbf{v})$ is the indicator function of the communication feasible set $\mathcal{C}=\Big\{\mathbf{v}\Big|\sum_{k\in\mathcal{K}}\mathbf{f}_k^T\mathbf{v}\leq\bar{R}_{c,k}(\mathbf{D}_p\mathbf{v})\Big\}$, and $g_r(\mathbf{u})$ is the indicator function of the radar feasible set $    \mathcal{R}=\Big\{\mathbf{u}\Big|\diag\big(\mathbf{D}_c\mathbf{u}\mathbf{u}^H\mathbf{D}_c^H+\sum_{k\in\mathcal{K}}\mathbf{D}_k\mathbf{u}\mathbf{u}^H\mathbf{D}_k^H\big)=\frac{P_t\mathbf{1}}{N}\Big\}$.

\normalsize
Finally, (\ref{admm_design}) is solved in an iterative updating manner as follows
{\small
\begin{equation}
    \begin{aligned}
        \mathbf{v}^{t+1}:=&\arg \min_{\mathbf{v}}\big(f_c(\mathbf{v}) +  g_c(\mathbf{v})+(\rho/2)||\mathbf{D}_p(\mathbf{v}-\mathbf{u}^t)+\mathbf{d}^t||_2^2\big),
    \end{aligned}
    \label{vupdate}
\end{equation}
\begin{equation}
    \begin{aligned}
       \mathbf{u}^{t+1}:=&\arg \min_{\mathbf{u}}\big(f(\mathbf{u}) + g_r( \mathbf{u})+(\rho/2)||\mathbf{D}_p(\mathbf{v}^{t+1}-\mathbf{u})+\mathbf{d}^t||_2^2\big),
    \end{aligned}
    \label{uupdate}
\end{equation}
\begin{equation}
    \begin{aligned}
        \mathbf{d}^{t+1}:=&\mathbf{d}^t+\mathbf{D}_p(\mathbf{v}^{t+1}-\mathbf{u}^{t+1}), 
    \end{aligned}
    \label{dupdate}
\end{equation}}
\normalsize
where $\mathbf{d} \in  \mathbb{C}^{N(K+1) \times 1}$ is the ADMM scaled dual variable and $\rho$ is the ADMM penalty parameter. The methods to perform the $\mathbf{v}$-update and the $\mathbf{u}$-update are explained next.

\subsection{AWSR Maximization Sub-problem}
The $\mathbf{v}$-update sub-problem in (\ref{vupdate}) is reformulated as follows
{\small
\begin{mini}|s|[2]
  {\bar{\mathbf{c}},\mathbf{P}}{-\sum_{k\in\mathcal{K}}\mu_k[\bar{C}_k+\bar{R}_k(\mathbf{P})]+\frac{\rho}{2}||\vect2(\mathbf{P})-\mathbf{D}_p\mathbf{u}^t+\mathbf{d}^t||_2^2,}{
  \label{awsr_optimization}}{}
  \addConstraint{\sum_{k'\in\mathcal{K}}\bar{C}_{k'}}{\leq\bar{R}_{c,k}(\mathbf{P})\;,\;\forall k\in\mathcal{K}}
  \addConstraint{\bar{\mathbf{c}}}{\geq \mathbf{0}}
  \addConstraint{\diag(\mathbf{P}\mathbf{P}^H)}{=\frac{P_t\mathbf{1}}{N}}
  \addConstraint{(\bar{C}_k + \bar{R}_k(\mathbf{P}))}{\geq \bar{R}_k^{th}\;,\;\forall k\in\mathcal{K}}.
\end{mini}}

\normalsize
Due to partial CSIT, the problem in (\ref{awsr_optimization}) is stochastic in nature. To solve it, the method proposed in \cite{joudeh} is adapted. Therefore, (\ref{awsr_optimization}) is first converted into a deterministic problem by employing the Sampled Average Approximation (SAA) method. Then, it is further transformed into a convex problem by applying the Weighted Minimum Mean Squared Error (WMMSE) approach and solved by using the Alternating Optimization (AO) algorithm.

\subsubsection{Sample Average Approximation}
For the current channel state estimate $\hat{\mathbf{H}}$ with conditional CSIT error distribution $f_{\text{H}|\hat{\text{H}}}(\mathbf{H}|\hat{\mathbf{H}})$, we generate a sample of $M$ i.i.d realizations indexed by the set $\mathcal{M}\triangleq\{1,\dots,M\}$ as follows:
\begin{equation*}
    \mathbb{H}^{(M)}\triangleq\{\mathbf{H}^{(M)}=\hat{\mathbf{H}}+\Tilde{\mathbf{H}}^{(M)}|\;\hat{\mathbf{H}},\;m \in \mathcal{M}\},
\end{equation*}
To compute $\mathbb{H}^{(M)}$, a set of $M$ normalized CSIT error realizations given by $\Tilde{\mathbb{H}}^{(M)}\triangleq\{\Tilde{\mathbf{H}}^{(M)}|\;m \in \mathcal{M}\}$ is first generated for a given $\sigma_e^2$. Then, for $\mathbf{H}^{(m)}$, the associated common and private rates can be defined as $R_{c,k}^{(m)}\triangleq R_{c,k}(\mathbf{H}^{(m)})$ and $R_k^{(m)}\triangleq R_k(\mathbf{H}^{(m)})$ respectively for user-$k$. These are then used to estimate the common and private ARs through their Sample Average Functions (SAFs), characterized by $\bar{R}_{c,k}\triangleq\frac{1}{M}\sum_{m=1}^MR_{c,k}^{(m)}$ and $\bar{R}_k\triangleq\frac{1}{M}\sum_{m=1}^MR_k^{(m)}$. The SAA of (\ref{awsr_optimization}) is then given by
{\small
\begin{mini}|s|[2]
  {\bar{\mathbf{c}},\mathbf{P}}{-\sum_{k\in\mathcal{K}}\mu_k[\bar{C}_k+\bar{R}_k^{(M)}]+\frac{\rho}{2}||\vect2(\mathbf{P})-\mathbf{D}_p\mathbf{u}^t+\mathbf{d}^t||_2^2,}{
  \label{saa_awsr_optimization}}{}
  \addConstraint{\sum_{k'\in\mathcal{K}}\bar{C}_{k'}}{\leq\bar{R}_{c,k}^{(M)},\quad}{\forall k\in\mathcal{K}}
  \addConstraint{\bar{\mathbf{c}}}{\geq \mathbf{0}}
  \addConstraint{\diag(\mathbf{P}\mathbf{P}^H)}{\leq\frac{P_t\mathbf{1}}{N}}
  \addConstraint{(\bar{C}_k + \bar{R}_k^{(M)})}{\geq R_k^{th}\;,\;\forall k\in\mathcal{K}},
\end{mini}}
\normalsize
where $\mathbf{P}$ is fixed for all $M$ channel realizations $\mathbf{H}^{(m)}$ and the average power per antenna constraint is relaxed from a non-convex equality to a convex inequality. From the strong Law of Large Numbers, the following relations are then stated:
\begin{equation}
    \begin{split}
        \lim_{M\rightarrow\infty}\bar{R}_{c,k}^{(M)}(\mathbf{P}) &= \bar{R}_{c,k}(\mathbf{P}),\;\text{almost surely}\;\forall\;\mathbf{P}\in\mathbb{P},\\
        \lim_{M\rightarrow\infty}\bar{R}_k^{(M)}(\mathbf{P}) &= \bar{R}_k(\mathbf{P}),\;\text{almost surely}\;\forall\;\mathbf{P}\in\mathbb{P},
    \end{split}
\end{equation}
where $\hat{\mathbb{P}} = \{\mathbf{P}\;|\;\diag(\mathbf{P}\mathbf{P}^H)\leq \frac{P_t\mathbf{1}}{N}\}$ is the feasible set of all possible precoders that fulfill the average power constraint. Therefore, as $M$ increases, the solutions to the SAA of the AWSR-maximization problem in (\ref{saa_awsr_optimization}) converge to those of the original stochastic problem in (\ref{awsr_optimization}).

\subsubsection{Weighted Minimum Mean Squared Error Approach}
The general WMMSE approach for 1-layer RSMA and $K$ communication users is then described. Additionally, it can be readily extended to more complex scenarios with multiple SIC layers by following the guidelines in \cite{rsma_lina}.  

User-$k$ first detects the common stream over the equalized signal $\hat{s}_{c,k}=g_{c,k}y_k$, where $g_{c,k}$ is the equalizer for the common stream. Then, it reconstructs the common stream using the decoded data and its corresponding channel vector $\mathbf{h}_k^H$ and performs error cancellation to detect the private stream over the equalized signal $\hat{s}_k=g_k(y_k-\mathbf{h}_k^H\mathbf{p}_c\hat{s}_{c,k})$, where $g_k$ is the equalizer for the private stream. 

The Mean Squared Error (MSE) of each stream is then defined as $\varepsilon\triangleq\mathbb{E}\big\{|s-\hat{s}|^2\big\}$ and can be calculated as
\begin{equation}
   \begin{split} \varepsilon_{c,k}&=|g_{c,k}|^2T_{c,k}-2\Re\big\{g_{c,k}\mathbf{h}_k^H\mathbf{p}_{c}\big\}+1,\\
   \varepsilon_k&=|g_k|^2T_k-2\Re\big\{g_k\mathbf{h}_k^H\mathbf{p}_k\big\}+1,
   \end{split}
   \label{mse_mmse}
\end{equation}
where $T_{c,k}\triangleq|\mathbf{h}_ k^H\mathbf{p}_{c}|^2+\sum_{j\in\mathcal{K}}|\mathbf{h}_ k^H\mathbf{p}_j|^2+1$ is the total received signal power and $T_k=T_{c,k}-|\mathbf{h}_ k^H\mathbf{p}_{c}|^2$. The optimum Minimum MSE (MMSE) equalizers can then be calculated by solving $\frac{\partial\varepsilon_{c,k}}{\partial g_{c,k}}=0$ and $\frac{\partial\varepsilon_k}{\partial g_k}=0$, which are given in closed form as  $g_{c,k}^{\text{MMSE}}=(\mathbf{p}_{c})^H\mathbf{h}_k\big(T_{c,k}\big)^{-1}$ and $g_k^{\text{MMSE}}=(\mathbf{p}_{k})^H\mathbf{h}_k\big(T_k\big)^{-1}$.

By using these into (\ref{mse_mmse}), the MMSEs can be achieved and are given by $\varepsilon_{c,k}^{\text{MMSE}}\triangleq\min_{g_{c,k}}\varepsilon_{c,k}=(T_{c,k})^{-1}I_{c,k}$ and $\varepsilon_k^{\text{MMSE}}\triangleq\min_{g_k}\varepsilon_k=(T_k)^{-1}I_k$, where $I_{c,k}=T_k$ is the power of the interference when decoding the common stream and $I_k=T_k-|\mathbf{h}_k^H\mathbf{p}_k|^2$ is the power of the interference when decoding the private stream. The SINRs of the common stream and the private stream of user-$k$ can then be expressed in terms of the MMSEs as $\gamma_{c,k}=1/\varepsilon_{c,k}^{\text{MMSE}}-1$ and $\gamma_k=1/\varepsilon_k^{\text{MMSE}}-1$. Consequently, the common stream rate $R_{c,k}$ and the private stream rate $R_k$ can be computed as $R_{c,k}=-\log_2\Big(\varepsilon_{c,k}^{\text{MMSE}}\Big)$ and $R_k=-\log_2\Big(\varepsilon_k^{\text{MMSE}}\Big)$.
        
The common and private streams at user-$k$ are assigned weights $\omega_{k}$ and $\omega_{c,k}$, respectively, and the augmented weighted MSEs (WMSEs) of user-$k$ are defined as 
\begin{equation}
    \begin{split}                \xi_{c,k}&=w_{c,k}\varepsilon_{c,k}-\log_2(w_{c,k}),\\   \xi_k&=w_k\varepsilon_k-\log_2(w_k).
    \end{split}
    \label{awmse_def}
\end{equation}
By solving $\frac{\partial\xi_{c,k}}{\partial g_{c,k}}=0$ and $\frac{\partial\xi_k}{\partial g_k}=0$, the optimum equalizers $(g_{c,k})^*=g_{c,k}^{\text{MMSE}}$ and $(g_k)^*=g_k^{\text{MMSE}}$ are derived and, by introducing them into (\ref{awmse_def}), the augmented WMMSEs can be expressed as $\xi_{c,k}\Big(g_{c,k}^{\text{MMSE}}\Big)=w_{c,k}(\varepsilon_{c,k}^{\text{MMSE}})-\log_2(w_{c,k})$ and $\xi_k\Big(g_k^{\text{MMSE}}\Big)=w_k(\varepsilon_k^{\text{MMSE}})-\log_2(w_k)$.

The optimum MMSE weights $(w_{c,k})^* = w_{c,k}^{\text{MMSE}}$ and $(w_{k})^* = w_{k}^{\text{MMSE}}$ can then be computed by solving $\frac{\partial \xi_{c,k}\big(g_{c,k}^{\text{MMSE}}\big)}{\partial w_{c,k}}=0$ and $\frac{\partial \xi_k\big(g_k^{\text{MMSE}}\big)}{\partial w_k}=0$ and are given by $w_{c,k}^{\text{MMSE}}\triangleq\Big(\varepsilon_{c,k}^{\text{MMSE}}\Big)^{-1}$ and $w_k^{\text{MMSE}}\triangleq\Big(\varepsilon_k^{\text{MMSE}}\Big)^{-1}$.
The Rate-WMMSE relationships are then found by using these into the augmented WMMSEs as follows
\begin{equation}
    \begin{split}                \xi_{c,k}^{\text{MMSE}}&\triangleq\min_{u_{c,k},g_{c,k}}\xi_{c,k}=1-R_{c,k},\\        \xi_k^{\text{MMSE}}&\triangleq\min_{u_k,g_k}\xi_k=1-R_k.
    \end{split}
    \label{rate_wmmse}
\end{equation}

In the presence of partial CSIT, the AR-WMMSE relationships can then be formulated for $\hat{\mathbf{H}}$. These are defined as the expected value of the Rate-WMMSE relationships over the conditional CSIT error distribution $f_{\text{H}|\hat{\text{H}}}(\mathbf{H}|\hat{\mathbf{H}})$ as
\begin{equation}
    \begin{split}
        \bar{\xi}_{c,k}^{\text{MMSE}}&\triangleq\mathbb{E}_{\text{H}|\hat{\text{H}}}\Big\{\min_{u_{c,k},g_{c,k}}\xi_{c,k}\;|\;\hat{\mathbf{H}}\Big\}=1-\bar{R}_{c,k},\\  \bar{\xi}_k^{\text{MMSE}}&\triangleq\mathbb{E}_{\text{H}|\hat{\text{H}}}\Big\{\min_{u_k,g_k}\xi_k\;|\;\hat{\mathbf{H}}\Big\}=1-\bar{R}_k.
    \end{split}
    \label{arwmmse}
\end{equation}
The Average WMSEs (AWMSEs) ($\bar{\xi}_{c,k}$, $\bar{\xi}_k$), along with the corresponding equalizers ($\bar{g}_{c,k}$, $\bar{g}_k$) and weights ($\bar{w}_{c,k}$, $\bar{w}_k$), can then be approximated by their SAFs to obtain a deterministic representation of the AR-WMMSE relationships in (\ref{arwmmse}) given by $\bar{\xi}_{c,k}^{(M)}\triangleq\frac{1}{M}\sum_{m=1}^M\xi_{c,k}^{(m)}$ and $ \bar{\xi}_{k}^{(M)}\triangleq\frac{1}{M}\sum_{m=1}^M\xi_{k}^{(m)}$,
where $(.)^{(m)}$ indicates that the parameter is associated to the $m^{th}$ channel realization in the set $\mathbb{H}^{(M)}$. Therefore, the set of  equalizers of the $M$ channel realizations is defined as $\mathbf{G}\triangleq\{\mathbf{g}_{c,k},\mathbf{g}_k\;|\;k\in\mathcal{K}\}$, where $\mathbf{g}_{c,k}\triangleq\{g_{c,k}^{(m)}\;|\;m\in\mathcal{M}\}$ and $\mathbf{g}_k\triangleq\{g_k^{(m)}\;|\;m\in\mathcal{M}\}$. In the same manner, the set of weights of the $M$ channel realizations is defined as $\mathbf{W}\triangleq\{\mathbf{w}_{c,k},\mathbf{w}_k\;|\;k\in\mathcal{K}\}$, where $\mathbf{w}_{c,k}\triangleq\{w_{c,k}^{(m)}\;|\;m\in\mathcal{M}\}$ and $\mathbf{w}_k\triangleq\{w_k^{(m)}\;|\;m\in\mathcal{M}\}$. In this way, (\ref{arwmmse}) can be represented as follows
\begin{equation}
    \begin{split}
        \big(\bar{\xi}_{c,k}\big)^{\text{MMSE}(M)}&\triangleq\min_{\mathbf{u}_{c,k},\mathbf{g}_{c,k}}\bar{\xi}_{c,k}^{(M)}=1-\bar{R}_{c,k}^{(M)},\\  \big(\bar{\xi}_k\big)^{\text{MMSE}(M)}&\triangleq\min_{\mathbf{u}_k,\mathbf{g}_k}\bar{\xi}_k^{(M)}=1-\bar{R}_k^{(M)}.
        \label{ar_wmse_rel}
    \end{split}
\end{equation}
{\normalsize
The set of optimum MMSE equalizers of the $M$ channel realizations is then defined as $\mathbf{G}^{\text{MMSE}}\triangleq\{\mathbf{g}_{c,k}^{\text{MMSE}},\mathbf{g}_k^{\text{MMSE}}\;|\;k\in\mathcal{K}\}$, where $\mathbf{g}_{c,k}^{\text{MMSE}}\triangleq\{g_{c,k}^{\text{MMSE}(m)}\;|\;m\in\mathcal{M}\}$ and $\mathbf{g}_k^{\text{MMSE}}\triangleq\{g_k^{\text{MMSE}(m)}\;|\;m\in\mathcal{M}\}$. In the same way, the set of of optimum MMSE weights of the $M$ channel realizations is defined as $\mathbf{W}^{\text{MMSE}}\triangleq\{\mathbf{w}_{c,k}^{\text{MMSE}},\mathbf{w}_k^{\text{MMSE}}\;|\;k\in\mathcal{K}\}$, where $\mathbf{w}_{c,k}^{\text{MMSE}}\triangleq\{w_{c,k}^{\text{MMSE}(m)}\;|\;m\in\mathcal{M}\}$ and $\mathbf{w}_k^{\text{MMSE}}\triangleq\{w_k^{\text{MMSE}(m)}\;|\;m\in\mathcal{M}\}$. The SAA-AWSR maximization problem in ($\ref{saa_awsr_optimization}$) can then be reformulated with the derived AR-WMSE relationships as follows}

\small{
\begin{mini}|s|[2]
  {\bar{\mathbf{x}},\mathbf{W},\mathbf{G},\mathbf{P}}{\sum_{k\in\mathcal{K}}\mu_k[\bar{X}_k+\bar{\xi}_k^{(M)}]+\frac{\rho}{2}||\vect2(\mathbf{P})-\mathbf{D}_p\mathbf{u}^t+\mathbf{d}^t||_2^2,}{
  \label{saa_awsr_optimization_arwmse}}{}
  \addConstraint{\bar{\xi}_{c,k}^{(M)}-1-\sum_{k'\in\mathcal{K}}\bar{X}_{k'}}{\leq 0,\quad}{\forall k\in\mathcal{K}}
  \addConstraint{\bar{\mathbf{x}}}{\leq \mathbf{0}}
  \addConstraint{\diag(\mathbf{P}\mathbf{P}^H)}{\leq\frac{P_t\mathbf{1}}{N}}
  \addConstraint{1-(\bar{X}_k+\bar{\xi}_k^{(M)})}{\geq R_k^{th}\;,\;\forall k\in\mathcal{K}},
\end{mini}
}\normalsize
where $\bar{\mathbf{x}}=[\bar{X}_{c,1},\dots,\bar{X}_{c,K}]^T=-[\bar{C}_1,\dots,\bar{C}_K]^T$ is a transformation of the previous variable $\bar{\mathbf{c}}$. 
\subsubsection{Alternating Optimization (AO) Algorithm}
The SAA-WMSE optimization problem in ($\ref{saa_awsr_optimization_arwmse}$) is non-convex if $\mathbf{W},\mathbf{G},\mathbf{P}$ are treated jointly. However, it is convex in each of $\mathbf{W},\mathbf{G}$ and $\mathbf{P}$, if the other two are fixed. Also, ($\mathbf{W},\mathbf{G}$) take the closed-form of the derived MMSE optimum solutions for a fixed $\mathbf{P}$. Therefore, the AO algorithm can be applied to solve ($\ref{saa_awsr_optimization_arwmse}$) by updating ($\mathbf{W},\mathbf{G}$) for a fixed $\mathbf{P}$ and then updating $\mathbf{P}$ with the recently updated ($\mathbf{W},\mathbf{G}$).

Thus, at the $n^{th}$ iteration of the AO algorithm,  $(\mathbf{W},\mathbf{G}) = (\mathbf{W}^{\text{MMSE}}(\mathbf{P}^{[n-1]}),\mathbf{G}^{\text{MMSE}}(\mathbf{P}^{[n-1]}))$  are first updated as the optimum MMSE weights and equalizers for the precoder matrix of the last iteration $\mathbf{P}^{[n-1]}$. The following parameters, used to express the AWMSEs, are also computed after updating $(\mathbf{W},\mathbf{G})$ \cite{joudeh}:

{\small
\begin{equation}
    \begin{split}                {t}_{c,k}^{(m)}=(w_{c,k}^{(m)}|g_{c,k}^{(m)}|^2)\quad&\text{and}\quad{t}_{k}^{(m)}=(w_k^{(m)}|g_k^{(m)}|^2),\\    
    {\mathbf{\Psi}}_{c,k}^{(m)}=(t_{c,k}^{(m)}\mathbf{h}_k^{(m)}\mathbf{h}_k^{(m)^H})\quad&\text{and}\quad{\mathbf{\Psi}}_{k}^{(m)}=(t_k^{(m)}\mathbf{h}_k^{(m)}\mathbf{h}_k^{(m)^H}),\\ 
    {\mathbf{f}}_{c,k}^{(m)}=(w_{c,k}^{(m)}\mathbf{h}_k^{(m)}g_{c,k}^{(m)^H})\quad&\text{and}\quad{\mathbf{f}}_{k}^{(m)}=(w_k^{(m)}\mathbf{h}_k^{(m)}g_k^{(m)^H}),\\
    {v}_{c,k}^{(m)}=(\log_2(w_{c,k}^{(m)}))\quad&\text{and}\quad{v}_k^{(m)}=(\log_2(w_k^{(m)})).
    \end{split}
    \label{safs_arwmse}
\end{equation}}

\normalsize
The SAFs $\bar{t}_{c,k}$, $\bar{t}_{k}$, $\bar{\mathbf{\Psi}}_{c,k}$, $\bar{\mathbf{\Psi}}_{k}$, $\bar{\mathbf{f}}_{c,k}$, $\bar{\mathbf{f}}_{k}$, $\bar{w}_{c,k}$, $\bar{w}_{k}$, $\bar{v}_{c,k}$ and $\bar{v}_{k}$ are then computed by calculating the average over the $M$ realizations. The AR-WMSE relationships in (\ref{saa_awsr_optimization_arwmse}) can be expanded, for every of the $M$ conditional channel realizations, by substituting the corresponding MSE as detailed in ($\ref{mse_mmse}$) into ($\ref{awmse_def}$), and the result into (\ref{ar_wmse_rel}). Then, these can be further expressed in terms of the derived SAFs in ($\ref{safs_arwmse}$) as shown in ($\ref{saa_awsr_optimization_arwmse_saf}$), which is the convex  Quadratically Constrained Quadratic Program (QCQP) to be solved to update $\mathbf{P}$.

{\small
\begin{mini}|s|[2]
  {\bar{\mathbf{x}},{\mathbf{P}}}{\sum_{k\in\mathcal{K}}\mu_k[\bar{X}_k+(\sum_{i=1}^K{\mathbf{p}}_i^H\bar{\mathbf{\Psi}}_k{\mathbf{p}}_i+\sigma_n^2\bar{t}_k-2\Re\{\bar{\mathbf{f}}_k^H{\mathbf{p}}_k\}+\bar{w}_k}{\label{saa_awsr_optimization_arwmse_saf}}{}\breakObjective{-\bar{v}_k)]+\frac{\rho}{2}||\vect2({\mathbf{P}})-\mathbf{D}_p\mathbf{u}^t+\mathbf{d}^t||_2^2,}
  \addConstraint{\makecell{({\mathbf{p}}_c^H\bar{\mathbf{\Psi}}_{c,k}{\mathbf{p}}_c+\sum_{i=1}^K{\mathbf{p}}_i^H\bar{\mathbf{\Psi}}_{c,k}{\mathbf{p}}_i+\sigma_n^2\bar{t}_{c,k}-2\Re\{\bar{\mathbf{f}}_{c,k}^H{\mathbf{p}}_c\}\\+\bar{w}_{c,k}-\bar{v}_{c,k})-1-\sum_{k'\in\mathcal{K}}\bar{X}_{k'}\leq 0,\quad\forall k\in\mathcal{K}}}{}{}
  \addConstraint{\bar{\mathbf{x}}}{\leq \mathbf{0}}
  \addConstraint{\diag({\mathbf{P}}{\mathbf{P}}^H)}{\leq\frac{P_t\mathbf{1}}{N}}
  \addConstraint{\makecell{1-[\bar{X}_k+(\sum_{i=1}^K{\mathbf{p}}_i^H\bar{\mathbf{\Psi}}_k{\mathbf{p}}_i+\sigma_n^2\bar{t}_k\\-2\Re\{\bar{\mathbf{f}}_k^H{\mathbf{p}}_k\}+\bar{w}_k-\bar{v}_k)]\geq R_k^{th},\quad\forall k\in\mathcal{K}.}}{}
\end{mini}
}

\normalsize
 Finally, the SAA AR-WMMSE-AO algorithm to solve the AWSR maximization problem is summarized in Algorithm $\ref{saa_ar_wmmse_ao_algorithm}$.
 {
\begin{algorithm}
\DontPrintSemicolon
  \KwInput{$n \leftarrow 0$, $\mathbf{P}^{[n]}$, $\text{AWSR}^{[n]}$;}
  \Repeat{$|\text{AWSR}^{[n]} - \text{AWSR}^{[n-1]}| \leq \epsilon$}
  {
  $n \leftarrow n+1$; \\
  $\mathbf{P}^{[n-1]} \leftarrow \mathbf{P};$\\
  $\mathbf{W}\leftarrow\mathbf{W}^{\text{MMSE}}(\mathbf{P}^{[n-1]});$\\
  $\mathbf{G}\leftarrow\mathbf{G}^{\text{MMSE}}(\mathbf{P}^{[n-1]});$\\
  Update the SAFs $\bar{\mathbf{\Psi}}_{c,k},\bar{\mathbf{\Psi}}_k,\bar{\mathbf{F}}_{c,k},\bar{\mathbf{F}}_k,\bar{t}_{c,k},\bar{t}_k,\bar{w}_{c,k},\bar{w}_k,\bar{v}_{c,k},\bar{v}_k\;,\;\forall k\in\mathcal{K}$;\\
  fix ($\mathbf{W},\mathbf{G}$) and use the computed SAFs to update ($\bar{\mathbf{x}},\mathbf{P}$) by solving the convex QCQP in ($\ref{saa_awsr_optimization_arwmse_saf}$);
  }
\caption{SAA AR-WMMSE-AO Algorithm}
\label{saa_ar_wmmse_ao_algorithm}
\end{algorithm}
}

\normalsize
\subsection{MSE Minimization Sub-problem}
The $\mathbf{u}$-update sub-problem in (\ref{uupdate}) is reformulated as follows

{\small
\begin{mini}|s|[2]
  {\alpha_u,\mathbf{p}_u}{\lambda\sum_{m=1}^M|\alpha_u \bm{P}_d(\theta_m)-\mathbf{a}_t^H(\theta_m)\big(\sum_{k=1}^{K+1}\mathbf{D}_{p,k}\mathbf{p}_u\mathbf{p}_u^H\mathbf{D}_{p,k}^H\big)}{
  \label{radar_optimization}}{}
  \breakObjective{\mathbf{a}_t(\theta_m)|^2+\frac{\rho}{2}||\mathbf{D}_p\mathbf{v}^{t+1}-\mathbf{p}_u+\mathbf{d}^t||_2^2,}
  \addConstraint{\diag\big(\sum_{k=1}^{K+1}\mathbf{D}_{p,k}\mathbf{p}_u\mathbf{p}_u^H\mathbf{D}_{p,k}^H\big)}{=\frac{P_t\mathbf{1}}{N}}
  \addConstraint{\alpha_u}{>0},
\end{mini}}

\normalsize
where $\alpha_u=u_1$ is the first entry of the optimization variable $\mathbf{u}$, $\mathbf{p}_u=[u_{K+2},u_{K+3},\dots,u_{(N+1)\times(K+1)}]^T \in \mathbb{C}^{N(k+1) \times 1}$, and $\mathbf{D}_{p,k} = [\mathbf{0}^{N \times (k-1)N}, \mathbf{I}_{N}, \mathbf{0}^{N \times (K+1-k)N}]$. Although ($\ref{radar_optimization}$) is originally non-convex, it can be transformed into a convex expression by employing Semi-Definite Relaxation (SDR) technique \cite{sdr}. 
Therefore, the new variable $\mathbf{R}_{\mathbf{p}_u}=\mathbf{p}_u\mathbf{p}_u^H$ is defined and the term $\big(\sum_{k=1}^{K+1}\mathbf{D}_{p,k}\mathbf{p}_u\mathbf{p}_u^H\mathbf{D}_{p,k}^H\big)$ is replaced with $\big(\sum_{k=1}^{K+1}\mathbf{D}_{p,k}\mathbf{R}_{\mathbf{p}_u}\mathbf{D}_{p,k}^H\big)$. Additionally, a new term $\hat{\mathbf{d}}_\mathbf{v}=\mathbf{D}_p\mathbf{v}^{t+1}+\mathbf{d}^t$ is introduced so that

\small{
\begin{equation*}   ||\mathbf{D}_p\mathbf{v}^{t+1}-\mathbf{p}_u+\mathbf{d}^t||_2^2=||\hat{\mathbf{d}}_\mathbf{v}||_2^2-2\Re\big\{\mathbf{p}_u^H\hat{\mathbf{d}}_\mathbf{v}\big\}+\text{Tr}(\mathbf{R}_{\mathbf{p}_u}).
\end{equation*}
}
\normalsize
The term $||\hat{\mathbf{d}}_\mathbf{v}||_2^2$ can be omitted in the optimization problem as it is a constant. 
By referring to \cite{sdr}, SDR can then be applied by Schur's complement. Let us define the matrix {\small $\mathbf{Q} = \begin{bmatrix}
        \mathbf{R}_{\mathbf{p}_u} & \mathbf{p}_u\\
        \mathbf{p}_u^H & 1
        \end{bmatrix}$},
\normalsize
which is a Hermitian semi-positive definite matrix ($\mathbf{Q}\succeq0$) and implies that $\mathbf{R}_{\mathbf{p}_u} \geq \mathbf{p}_u\mathbf{p}_u^H$ and  $\text{Tr}(\mathbf{Q})=\text{Tr}(\mathbf{R}_{\mathbf{p}_u})+1$.
This is the key convex relaxation that allows ($\ref{radar_optimization}$) to turn into the following convex optimization problem

\small{
\begin{mini}|s|[2]
  {\alpha_u,\mathbf{p}_u,\mathbf{R}_{\mathbf{p}_u}}{\lambda\sum_{m=1}^M|\alpha_u \bm{P}_d(\theta_m)-\mathbf{a}_t^H(\theta_m)\big(\sum_{k=1}^{K+1}\mathbf{D}_{p,k}\mathbf{R}_{\mathbf{p}_u}\mathbf{D}_{p,k}^H\big)}{\label{radar_optimization_convex}}{}
  \breakObjective{\mathbf{a}_t(\theta_m)|^2+\frac{\rho}{2}\big(\text{Tr}(\mathbf{Q})-2\Re\big\{\mathbf{p}_u^H\hat{\mathbf{d}}_\mathbf{v}\big\}-1\big),}
  \addConstraint{\diag\big(\sum_{k=1}^{K+1}\mathbf{D}_{p,k}\mathbf{R}_{\mathbf{p}_u}\mathbf{D}_{p,k}^H\big)}{=\frac{P_t\mathbf{1}}{N}}
  \addConstraint{\mathbf{Q}}{\succeq0}
  \addConstraint{\alpha_u}{>0.}
\end{mini}
}

\normalsize
\subsection{ADMM Algorithm}
The ADMM-based optimization algorithm is summarized in Algorithm $\ref{radcom_admm_algorithm_imperfect}$. The process is repeated iteratively until the primal residual $\mathbf{r}^{t+1}$ and the dual residual $\mathbf{q}^{t+1}$ of the ADMM algorithm converge to a value below a predefined threshold $\nu$.
\begin{algorithm}
\DontPrintSemicolon
  \KwInput{$t \leftarrow 0$, $\mathbf{v}^t$, $\mathbf{u}^t$,
  $\mathbf{d}^t$;}
  \Repeat{$||\mathbf{r}^{t+1}||_2 \leq \nu$ and $||\mathbf{q}^{t+1}||_2 \leq \nu$}
  {
  $\mathbf{v}^{t+1}\leftarrow\arg \min_{\mathbf{v}}\big(f_c(\mathbf{v}) + g_c(\mathbf{v})+(\rho/2)||\mathbf{D}_p(\mathbf{v}-\mathbf{u}^t)+\mathbf{d}^t||_2^2\big) $ using SAA AR-WMMSE-AO;\\
  $\mathbf{u}^{t+1}\leftarrow\arg \min_{\mathbf{u}}\big(f(\mathbf{u}) + g( \mathbf{u})+(\rho/2)||\mathbf{D}_p(\mathbf{v}^{t+1}-\mathbf{u})+\mathbf{d}^t||_2^2\big)$ using SDR;\\
  $\mathbf{d}^{t+1}\leftarrow\mathbf{d}^t+\mathbf{D}_p(\mathbf{v}^{t+1}-\mathbf{u}^{t+1});$\\
  $\mathbf{r}^{t+1}=\mathbf{D}_p(\mathbf{v}^{t+1}-\mathbf{u}^{t+1});$\\
  $\mathbf{q}^{t+1}=\mathbf{D}_p(\mathbf{u}^{t+1}-\mathbf{u}^{t});$\\
  $t \leftarrow t+1$; 
  }
\caption{ADMM-based DFRC optimization algorithm with partial CSIT}
\label{radcom_admm_algorithm_imperfect}
\end{algorithm}

\section{Simulation Results}
\label{numerical_section}
In this section, the proposed DFRC is evaluated in terms of the Ergodic Weighted Sum-Rate (EWSR) and Root Mean Squared Error (RMSE) trade-off, and average precoder power distribution. The effectiveness of the ADMM-based DFRC optimization algorithm is then also validated using the conventional radar metrics CRB and RMI. Finally, LLS results are discussed, which further demonstrate the superiority of an RSMA-based DFRC over employing SDMA or NOMA in the DFRC design as indicated in subsection \ref{baseline_subsection}. Results for perfect CSIT optimization as proposed
in \cite{chengcheng} are also included in order to highlight the robustness and flexibility of the RSMA-based DFRC as the CSIT quality degrades.

\subsection{Simulation Parameters}
MATLAB is used to run all simulations, with the optimization problems being solved using the CVX toolbox \cite{cvx_web,cvx_notes}. The average of the optimization results for 200 different channel realizations are used to generate all plots.

It is assumed that the DFRC employs $N=8$ transmit antennas arranged in a ULA structure, with the spacing between elements being $\delta = 0.5$ wavelengths, to serve $K=4$ communication users and track a single radar target located in the azimuth angle range $[-8, 8]^{\circ}$. The total available transmit power is $P_t=20 \text{ dBm}$ and the noise power at each user is $\sigma_n^2=0 \text{ dBm}$. The precoder matrix $\mathbf{P}$ is initialized following the Maximum Ratio Transmission (MRT) and Singular Value Decomposition (SVD) technique introduced in \cite{joudeh}. The desired radar beampattern $\bm{P}_d$ to approximate during optimization is obtained following the method described in \cite{stoica_radar}, and the beampattern scaling factor is initialized as $\alpha=1$. Finally, the initial elements in $\mathbf{d}$ are drawn from the distribution $\mathcal{CN}(0,1)$ and the QoS rate constraint for each communication user is 0.1 bps/Hz.

\begin{figure}[t!]
		\centering
        \includegraphics[width=0.49\textwidth]{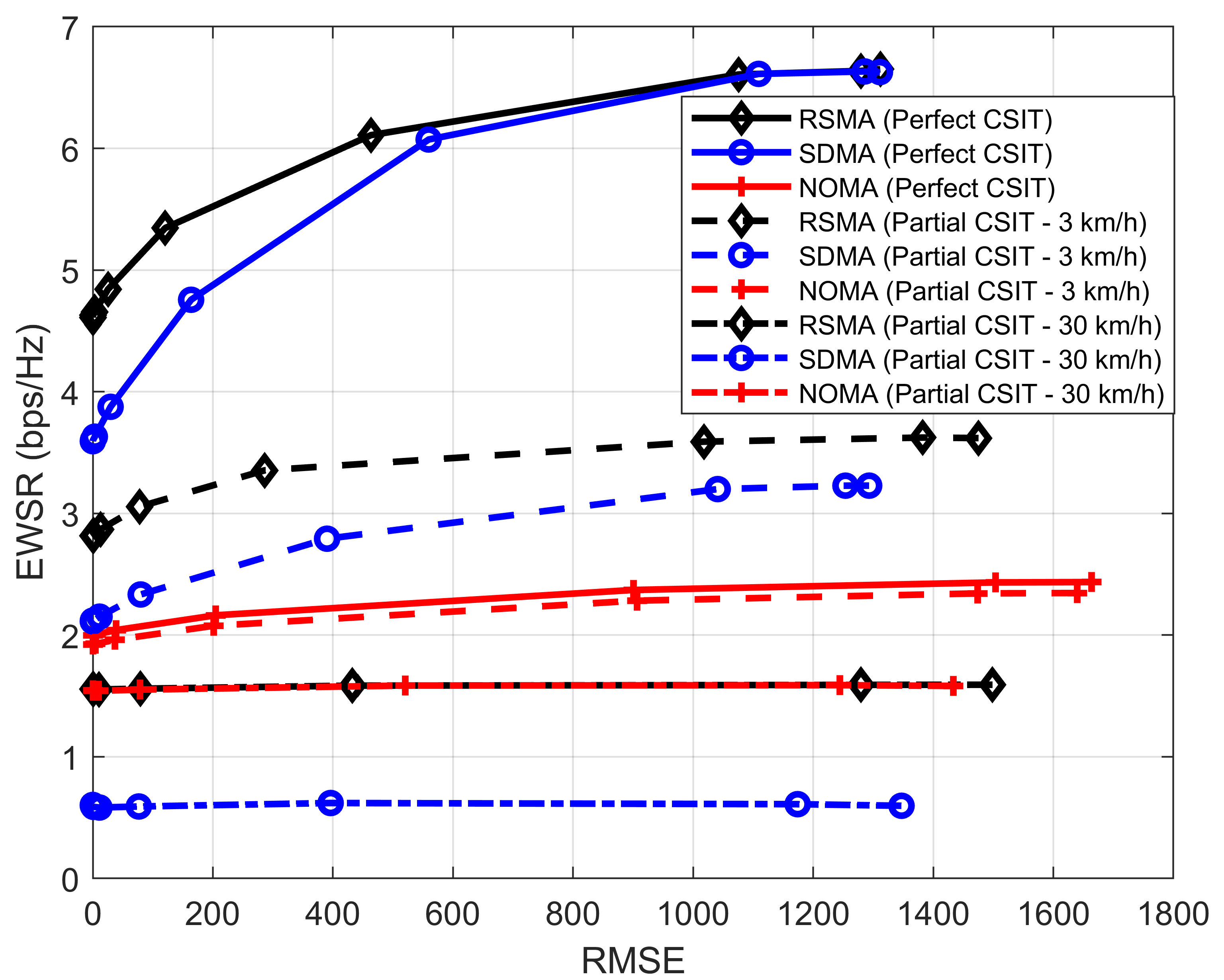}
		\caption{EWSR vs. RMSE. }
		\label{fig:ewsr_erbse}
\end{figure} 
\begin{figure}[t!]
		\centering
        \includegraphics[width=0.49\textwidth]{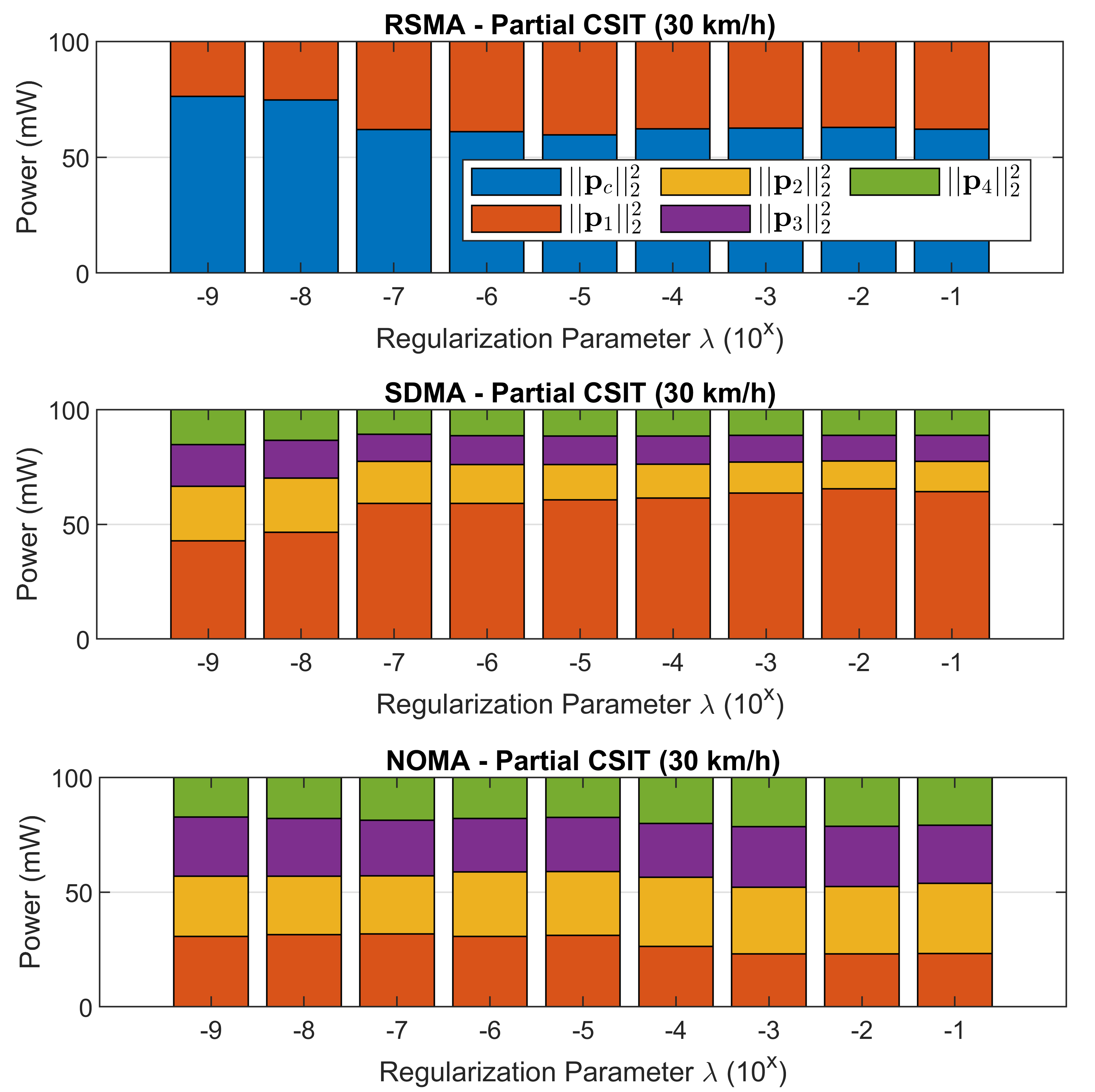}
		\caption{DFRC precoder power distribution.}
		\label{fig:pow_30}
\end{figure} 
\subsection{Communication Users Mobility Model for Partial CSIT}
In order to consider a practical scenario, we assume that partial CSIT is caused by the mobility of the communication users and delays in the CSI acquisition process to the DFRC. This problem is also known as channel aging. To model this, we assume in (\ref{csi_eq}) that $\sqrt{\sigma_k^2-\sigma_e^2}=J_0(2\pi f_DT)$ is the time correlation coefficient that follows Jakes' model \cite{jakes}, where $J_0(.)$ is the Bessel function of order 0, $T$ is the channel sampling interval, $f_D$ = $\frac{vf_c}{c}$ is the maximum Doppler frequency for a given user velocity $v$, carrier frequency $f_c$ , and $c \approx 3\times10^8$ m/s is the speed of the EM waves in air. Also, the matrix $\mathbf{N}_{m}$ has i.i.d entries drawn from the distribution $\mathcal{CN}(0,1)$. Due to latency in the CSI acquisition process, the DFRC only knows $\hat{\mathbf{H}}_m=\mathbf{H}_{m-1}$ at time instant $m$. We highlight then that, effectively, $\sigma_e^2 = \sqrt{\sigma_k^2-(J_0(2\pi f_DT))^2}$ is used to generate the set $\mathbb{H}^{(M)}$ for the $\mathbf{v}$-update sub-problem in (\ref{awsr_optimization}). We then define two cases with different degrees of CSIT imperfections, assuming that $T = 10\text{ ms}$. The first assumes that the communication users move at a constant velocity of $v = 3$ km/h ($\sigma_e^2=0.417$), and in the second, at $v = 30$ km/h ($\sigma_e^2=0.984$) \cite{onur_parameters}.

\subsection{EWSR vs. RMSE Trade-off and Average Precoder Power Distribution}

The computed EWSR vs. RMSE trade-off curves are plotted in Fig. \ref{fig:ewsr_erbse}, where the rightmost points correspond to prioritizing communications ($\lambda = 10^{-9}$), and the leftmost points, to prioritizing radar functions ($\lambda = 10^{-1}$). The average precoder power allocation of the three DFRCs for $v=30$ km/h, over the 200 different channel realizations as $\lambda$ varies, are also shown in Fig. \ref{fig:pow_30}. User ordering is done according to the instantaneous channel gain of each user, i.e. $||\mathbf{h}_k||_2^2,\forall\;k\in\mathcal{K}$, in descending order. 

From Fig. \ref{fig:ewsr_erbse}, it can be noticed that the RSMA-based DFRC achieves the best performance regardless of the CSIT quality. With perfect CSIT, it is seen that the RSMA-based SDMA-based DFRCs achieve similar EWSR when prioritizing communications, but as the priority is shifted to radar, the RSMA-based DFRC outperforms the SDMA-based DFRC. As detailed in \cite{chengcheng}, this is possible due to the unique common stream in RSMA, which plays a crucial role to generate a directional radar beampattern. As the CSIT quality worsens, the role of the common stream becomes more important for both radar and communications. In the high mobility setting with $v=30$ km/h, it can be observed from Fig. \ref{fig:pow_30} that the power is distributed only between $\mathbf{p}_c$ and $\mathbf{p}_1$. This is done in order to comply with the QoS rate constraints, as user-2, user-3 and user-4 are served only through the common stream. Since user-1 experiences the best channel conditions, its private rate can be maximized without interference of any other private streams after removing the common stream interference in the SIC process.
\begin{figure*}[t!]
\centering
\begin{minipage}{.47\textwidth}
   \centering
       \includegraphics[width=\textwidth]{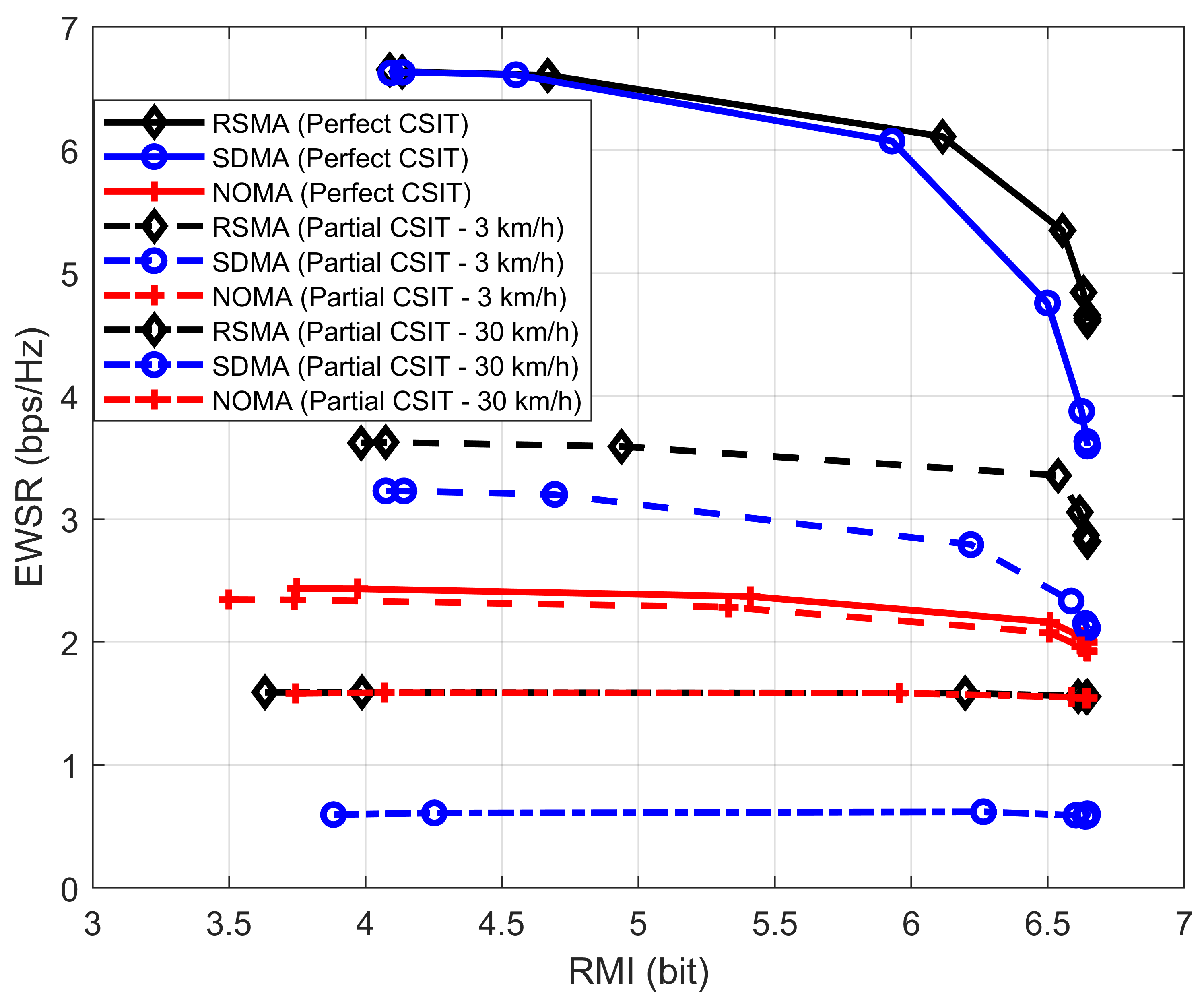}
		\caption{EWSR vs. RMI.}
		\label{fig:ewsr-rmi}
\end{minipage}\hfill
\begin{minipage}{.48\textwidth}
   \centering
       \includegraphics[width=\textwidth]{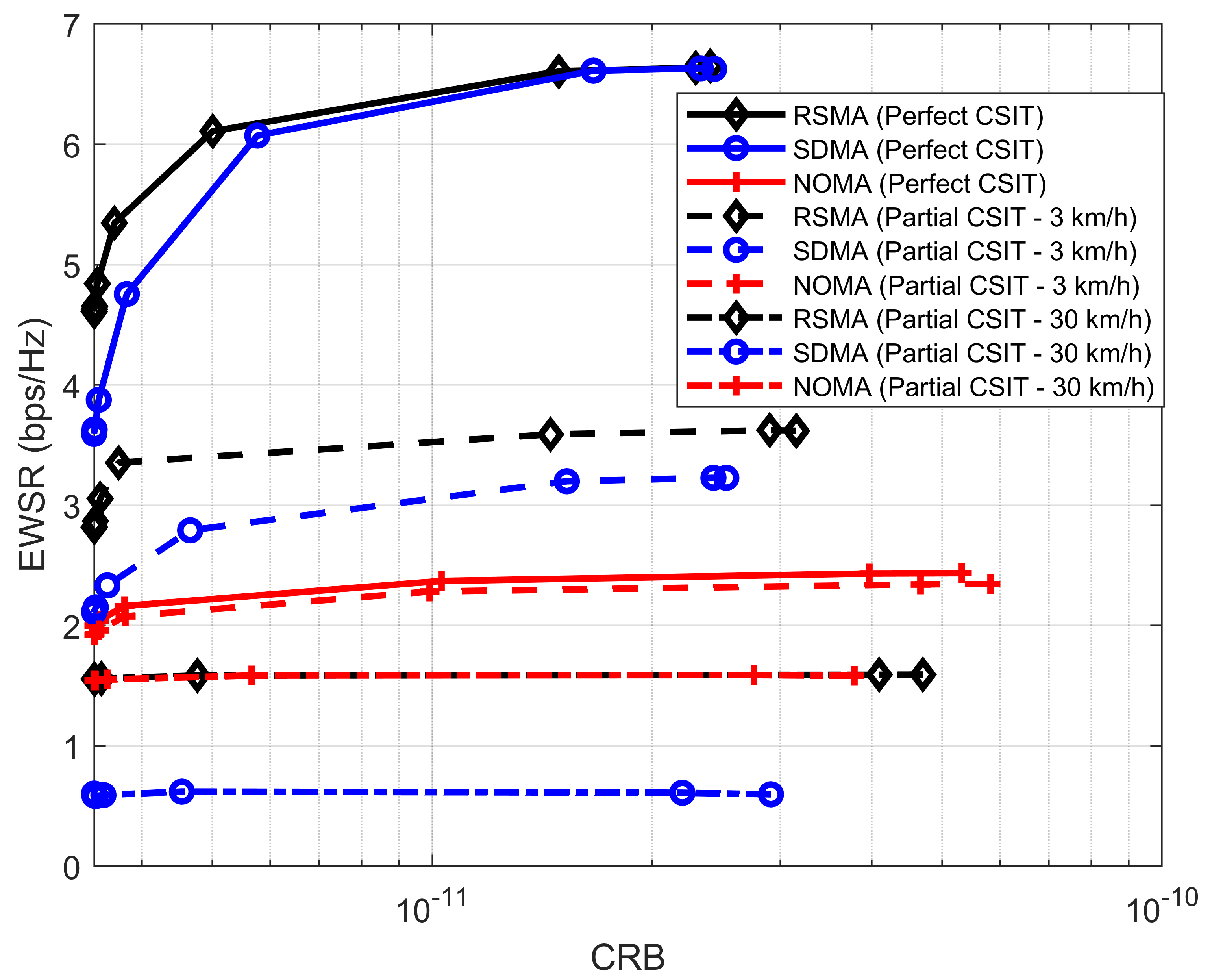}
		\caption{EWSR vs. CRB.}
		\label{fig:ewsr-crb}
\end{minipage}
\end{figure*}

The sub-optimality of the SDMA-based and NOMA-based DFRCs can also be observed from Fig. \ref{fig:ewsr_erbse}. When prioritizing communications, the SDMA-based DFRC relies greatly on perfect CSIT to spatially separate the user streams, minimize the MUI and achieve full spatial multiplexing gain \cite{rsma_lina}. Thus, the MUI due to CSIT errors causes a severe performance loss which increases with $\sigma_e^2$. When switching to radar instead, the SDMA-based DFRC increases the power allocated to $\mathbf{p}_1$ to approximate $\bm{P}_d$. For instance, consider the high mobility scenario with $v=30$ km/h. It can be seen from Fig. \ref{fig:pow_30} that the power of $\mathbf{p}_1$ increases by 20 dBm when fully shifting to radar priority. However, when compared to the RSMA-based and NOMA-based DFRCs, it is seen that the EWSR of these is twice as large as the EWSR achieved by the SDMA-based DFRC. Since enough power needs to be allocated to the precoders of the other users to comply with the QoS rate constraints, user-1 incurs MUI and its rate cannot be increased even though it experiences the best channel conditions. 

The NOMA-based DFRC achieves the poorest performance except for $v=30$ km/h where it performs similarly to the RSMA-based DFRC. This performance loss is due to the inefficient use of SIC layers in multi-antenna NOMA, which lowers the sum DoF compared to RSMA and SDMA \cite{noma_clerckx}. Specifically, the rate of user-1 can only be maximized at the expense of minimizing the rates of all other users to levels close to the QoS rate limit. When optimizing the precoders for radar when $v=30$ km/h, it is observed from Fig. \ref{fig:pow_30} that the NOMA-based DFRC decreases the power allocated to the precoder of user-1 as, otherwise, larger MUI levels would be experienced by the other users, which would further decrease the minimum rate supported in the SIC process and the overall EWSR as a result. 

\subsection{EWSR Trade-off with Conventional Radar Sensing Metrics}
Using the optimized precoders, the RMI and CRB can be calculated. The EWSR trade-offs with these conventional radar sensing metrics are then shown in Fig. \ref{fig:ewsr-rmi} and \ref{fig:ewsr-crb} for RMI and CRB, respectively. The path loss $h_0$ in (\ref{path_loss_eq}) is calculated using the standard Friis' free space path loss model \cite{friis}, assuming that the radar target is located 50 m away from the DFRC. Also, the noise power at the receiver antenna of the DFRC is -150 dBm. Finally, the velocity of the radar target is assumed to be $v_0=3$ m/s.

Overall, these trade-off curves present highly similar trends to the EWSR vs. RMSE trade-off in the previous subsection. This is expected as the impact of appropriate beampattern shaping of the DFRC with the RMI and CRB was described in Subsection \ref{other_metric_disc}. Small differences with the EWSR vs. RMSE trade-off are only observed when communications are prioritized as in this region the precoders are not optimized to generate the desired radar beampattern. Thus, the beampattern gain in the $0^{\circ}$ azimuth direction may fluctuate largely which directly affects the RMI and CRB.
\begin{figure*}[t!]
		\centering
        \includegraphics[width=\textwidth]{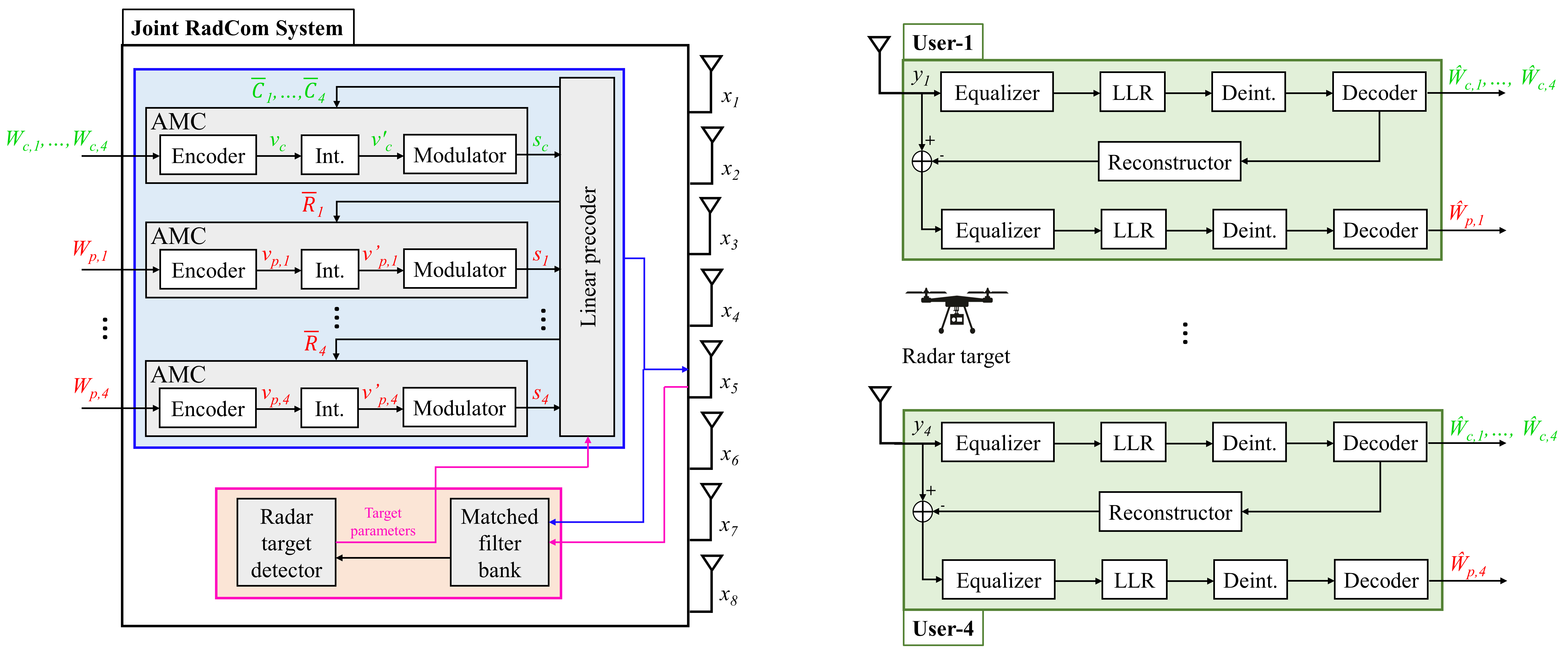}
		\caption{1-Layer RSMA transmitter and receiver structures used in LLS \cite{onur_lls}.}
		\label{fig:lls_arch}
\end{figure*}       

\begin{figure}[t!]
		\centering
        \includegraphics[width=0.49\textwidth]{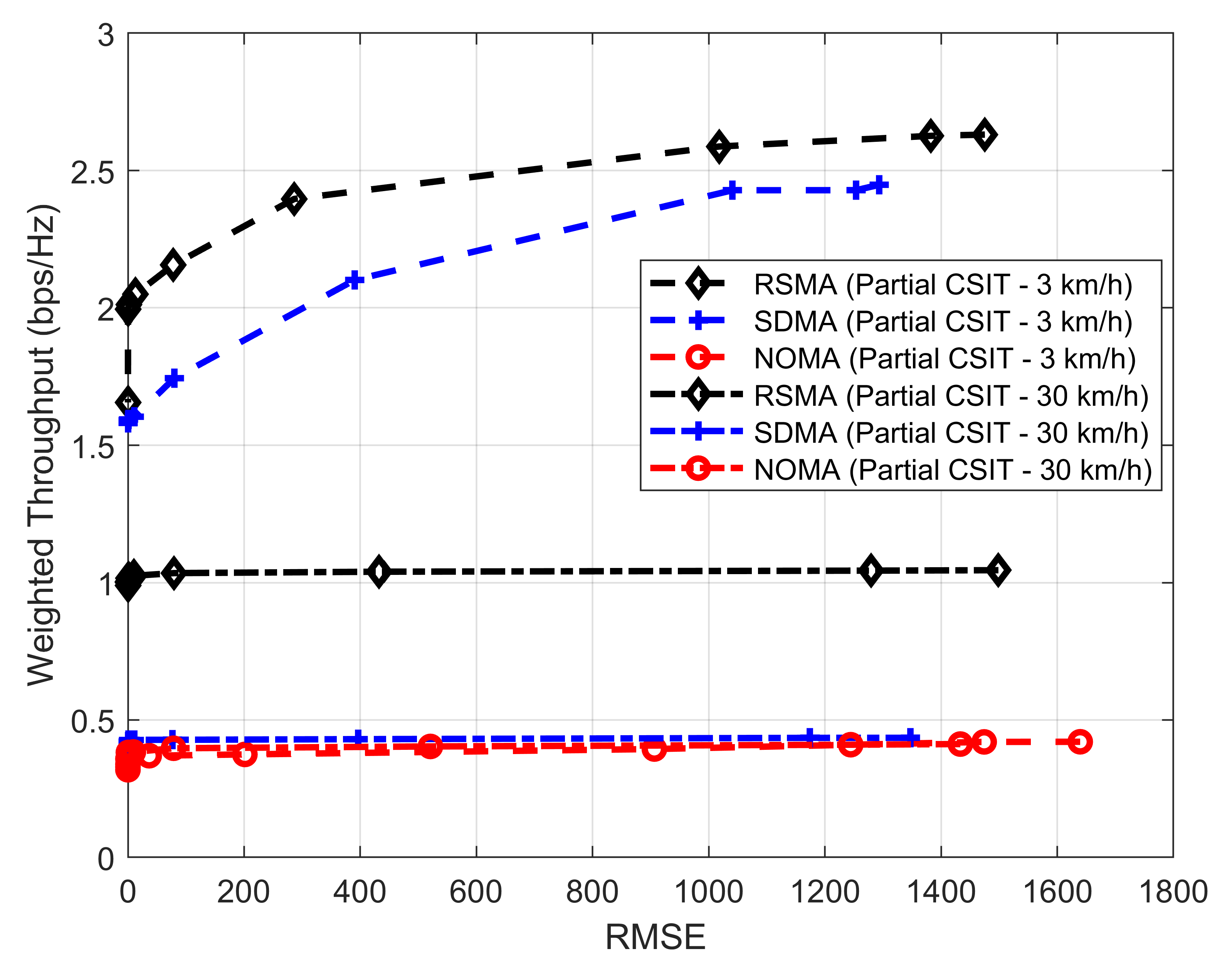}
		\caption{Weighted Throughput vs. RMSEs.}
		\label{fig:throughput_rbse}
\end{figure} 
\subsection{Link-Level Simulations}
The proposed DFRC is then further evaluated through LLS. This is performed by adapting the 1-Layer RSMA transceiver architecture described in \cite{onur_lls}, as depicted in Fig. \ref{fig:lls_arch}. The LLS framework features finite alphabet modulation schemes (QPSK, 16-QAM, 64-QAM or 256-QAM), finite-length channel coding (polar codes \cite{polar_code}) and an Adaptive Modulation and Coding (AMC) algorithm, where the latter selects an appropriate modulation-coding rate pair based on the achievable transmit rate calculation from the precoder optimization. For each of the 200 channel realizations, it is considered that, once precoder optimization is performed with partial CSIT, the transmit rates are calculated assuming the DFRC knows the real channel in order to always select the best possible modulation-coding rate pair.

A new performance metric for communications is then introduced, denominated Weighted Throughput and defined as
\begin{equation}
    \text{Weighted Throughput [bps/Hz]} = \frac{\sum_l \sum_k \mu_k D_{s,k}^{(l)}}{\sum_l S^{(l)}},
\end{equation}
where $S^{(l)}$ denotes the modulated block length in the $l$-th Monte Carlo channel realization, and $D_{s,k}^{(l)}$ denotes the number of received bits by user-$k$ in the common stream (considering only its intended part of the common message) and private stream when there are no decoding errors in a given block. In all simulations, $S^{(l)} = 256$ is used. The Weighted Throughput vs. RMSE trade-off is then shown in Fig.\ref{fig:throughput_rbse}.

Although the degradation due to the practical considerations in modulation and coding are seen when compared with the theoretical Shannon Bounds in Fig. \ref{fig:ewsr_erbse}, it is nevertheless evident that the RSMA-based DFRC still provides the best trade-off in both low and high mobility scenarios. Thus, the superiority of employing RSMA to empower the proposed DFRC is confirmed. Finally, one important observation is made about the NOMA-based DFRC as it can be immediately noticed that it achieves the worst performance in both mobility settings: due to the multiple SIC layers, there may be decoding errors to occur before each user decodes its own message. Consider user-1, which is expected to be the main contributor to the total throughput. As decoding errors occur somewhere along the 3 SIC layers it employs, then it does not contribute to the Weighted Throughput and the NOMA-based DFRC largely underperforms.

\section{Conclusion}
\label{conclusion_section}
We investigate a DFRC system to simultaneously maximize the AWSR and minimize the MSE against a desired highly-directional transmit beampattern with partial CSIT. As it is necessary to mitigate the MUI caused by CSIT inaccuracies and by the forced directional radar beampattern generation, it is proposed that the DFRC employs RSMA as the communications scheme to partially decode MUI and partially treat it as noise. An ADMM-based optimization algorithm is then introduced to optimize the DFRC precoders to achieve this. Simulation results reveal that this proves to be a more robust and flexible approach to comply with rate QoS constraints than the SDMA-based and NOMA-based DFRCs, and to further minimize the DFRC beampattern MSE. It is also demonstrated that minimizing the radar beampattern MSE translates into maximizing the RMI and minimizing the CRB of the DFRC. The benefits and superiority of employing RSMA in the DFRC are further demonstrated through LLS even with the limitations induced by finite length coding and finite modulation schemes. Future work could focus on optimizing the modulation and coding schemes, and receiver architectures in order to minimize the gap between the throughput levels obtained in the LLS and the sum-rate of the theoretical Shannon Bounds so that the proposed DFRC can be employed in a practical deployment.

\end{document}